\documentclass[11pt,english]{article}
\usepackage[T1]{fontenc}
\usepackage[latin9]{inputenc}
\usepackage{verbatim}
\usepackage{mathtools}
\usepackage{amsmath}
\usepackage{amssymb}
\usepackage{stackrel}
\usepackage{graphicx}
\usepackage[numbers]{natbib}

\makeatletter

\providecommand{\tabularnewline}{\\}

\pdfoutput=1
\usepackage[T1]{fontenc}
\usepackage[latin9]{inputenc}
\usepackage{color}
\usepackage{float}
\usepackage{graphicx}
\usepackage{graphics}
\usepackage{esint}
\usepackage{enumerate}

\makeatletter

\providecommand{\tabularnewline}{\\}

\usepackage{latexsym}
\usepackage{epsfig}
\usepackage{graphicx}
\usepackage[colorlinks,linkcolor=blue]{hyperref}
\allowdisplaybreaks 

\usepackage{babel}
\usepackage{tcolorbox}

\usepackage{tikz}

\makeatother

\usepackage{babel}
\begin{document}
{}~ \hfill\vbox{\hbox{CTP-SCU/2026007}}\break
\vskip 2.5cm
\centerline{\Large \bf String charge density/bit threads correspondence} 
\vspace*{1.0ex}

\centerline{\Large \bf and its S-duality in type IIB supergravity} 

\vspace*{11.0ex}
\centerline{\large  Houwen Wu and Shuxuan Ying}
\vspace*{7.0ex}
\vspace*{3.0ex}

\centerline{\large \it College of Physics}
\centerline{\large \it Sichuan University}
\centerline{\large \it Chengdu, 610065, China} \vspace*{1.0ex}

\vspace*{3.0ex}
\centerline{\large \it Department of Physics}
\centerline{\large \it Chongqing University}
\centerline{\large \it Chongqing, 401331, China} \vspace*{1.0ex}
\vspace*{3.0ex}

\centerline{iverwu@scu.edu.cn, ysxuan@cqu.edu.cn}
\vspace*{8.0ex}
\centerline{\bf Abstract} \bigskip \smallskip
We identify the spatial projection of a conserved string charge density with a bit-thread flow. This correspondence gives bit threads a string worldsheet description and allows us to compute black hole entropy from the maximal entropy flux of a worldsheet-derived current. We first reproduce the BTZ entropy and then extend the construction to the ten-dimensional F1-NS5-P and D1-D5-P systems. Under Type IIB S-duality, the F1 current maps to the D1 current. Although the local currents, metrics, and flow norms change, the maximal physical flux remains unchanged and gives the same entropy. This provides a nontrivial check of the correspondence. Finally, we propose a D1-D5 boundary-matching conjecture. In the coarse-grained picture, each string-charge line defines an entropy-carrying tube. The boundary CFT entanglement entropy and the bulk black hole entropy both fix the product of the number of tubes and the maximum entropy carried by each tube. For a fixed tube discretization, they assign the same capacity to each tube. They therefore describe the same conserved maximal entropy flux, providing an information-flow interpretation of the equality between entanglement entropy and black hole entropy.

\vfill 
\eject
\baselineskip=16pt
\vspace*{10.0ex}
\tableofcontents

\section{Introduction}

The Bekenstein--Hawking entropy implies that a black hole possesses
a large number of microscopic degrees of freedom whose logarithmic
degeneracy is measured geometrically by the area of its horizon. Understanding
the nature of these degrees of freedom, and how they encode information
during black hole evaporation remains one of the central problems
in quantum gravity. A major advance was provided by the Ryu--Takayanagi
(RT) prescription and its covariant extension, which relate the von
Neumann entropy of a boundary region $A$ in a holographic conformal
field theory to the area of a bulk codimension-two extremal surface
$\gamma_{A}$ homologous to $A$ \cite{Ryu:2006bv,Ryu:2006ef,Hubeny:2007xt}.
In the classical limit, 

\begin{equation}
S\left(A\right)=\frac{\mathrm{Area}\left(\gamma_{A}\right)}{4G_{N}}.
\end{equation}

\noindent For AdS$_{3}$/CFT$_{2}$, the surface $\gamma_{A}$ is
a bulk geodesic, and the area is its geodesic length. Including bulk
quantum effects leads to the quantum extremal surface prescription
\cite{Faulkner:2013ana,Engelhardt:2014gca},

\begin{equation}
S\left(A\right)=\underset{X\sim A}{\mathrm{min}}\underset{X}{\mathrm{ext}}\left[\frac{\mathrm{Area}\left(X\right)}{4G_{N}}+S_{\mathrm{bulk}}\left(\Sigma_{X}\right)\right],
\end{equation}

\noindent where $S_{\mathrm{bulk}}\left(\Sigma_{X}\right)$ is the
bulk entanglement entropy across $X$. Applied to evaporating black
holes, this prescription gives rise to entanglement islands and reproduces
the expected Page curve in controlled semiclassical models \cite{Penington:2019npb,Almheiri:2019psf}.
These developments provide strong evidence that semiclassical gravity
correctly captures the unitary evolution of the fine-grained entropy.
However, a complete microscopic identification of the degrees of freedom
whose entanglement accounts for the entropy of Hawking radiation remains
unavailable, posing a key unresolved aspect of the information paradox
\cite{Almheiri:2020cfm}.

A string perspective was proposed by Susskind and Uglum \cite{Susskind:1994sm}.
They argued that black-hole entropy could be described by closed-string
sphere diagrams that intersect the black-hole horizon twice. This
picture can also be interpreted from the open string perspective,
in which the string endpoints are fixed on the horizon. In this framework,
a slice of the closed string intersected by the horizon appears as
an open string to a Rindler observer outside the event horizon, offering
a statistical interpretation of entanglement entropy. The Susskind--Uglum
proposal thus suggests that both black hole entropy and entanglement
entropy can be computed in string theory, potentially allowing the
identification of their underlying quantum states. Therefore, completing
the Susskind--Uglum picture may provide a possible way to understand
the microscopic structure of black holes and the quantum state corresponding
to the entanglement entropy of Hawking radiation. 

More recently, Ahmadain and Wall placed the closed string side of
this proposal on a firmer footing \cite{Ahmadain:2022tew,Ahmadain:2022eso,Ahmadain:2024hdp}.
Using an off-shell formulation of closed string theory, they derived
the leading-order closed string effective action from sphere diagrams
and showed how the Bekenstein--Hawking term can be recovered from
the renormalization-group flow of the nonlinear sigma model on conical
target spaces. Their analysis reproduces the Bekenstein--Hawking
entropy at leading order in $\alpha^{\prime}$. The open string picture
is more subtle because the replica geometry is off shell and contains
a conical singularity, so the corresponding worldsheet theory is not
an ordinary conformal background. This difficulty motivates the search
for a formulation that relates string charge, entanglement entropy,
and extremal surfaces without requiring a direct quantization of strings
on the singular replica geometry. Several related studies have investigated
black hole entropy from the perspective of string worldsheet theory
\cite{He:2014gva,Brustein:2022wiq,Halder:2023adw,Mori:2025qoh,Ahmadain:2025pox,Jorstad:2026jlg}. 

In our recent work \cite{Wu:2025qwc}, we proposed such a relation
in the context of AdS$_{3}$/CFT$_{2}$. We began with the von Neumann
entanglement entropy $S_{\mathrm{vN}}$ of a CFT$_{2}$. Using the
RT prescription, the geodesic distance $L\left(X,X^{\prime}\right)$
between the corresponding bulk points satisfies

\begin{equation}
L\left(X,X^{\prime}\right)=4G_{N}^{\left(3\right)}S_{\mathrm{vN}},
\end{equation}

\noindent Consequently, Synge\textquoteright s world function \cite{Synge:1960}
can be reconstructed from the entanglement entropy as

\begin{equation}
\Omega\left(X,X^{\prime}\right)=\frac{1}{2}L\left(X,X^{\prime}\right)^{2}=8\left(G_{N}^{\left(3\right)}\right)^{2}S_{\mathrm{vN}}^{2},
\end{equation}

\noindent The bulk points $X$ and $X^{\prime}$ are determined by
the boundary entangling data and can therefore be parameterized by
two variables $\left(\tau,\sigma\right)$ of boundary CFT. Varying
the boundary intervals generates a two-parameter family of geodesics
that sweeps out an effective two-dimensional surface,

\begin{equation}
X^{\mu}=X^{\mu}\left(\tau,\sigma\right).
\end{equation}

\noindent This observation led us to construct an action directly
from $\Omega\left(X,X^{\prime}\right)$, with $\left(\tau,\sigma\right)$
interpreted as coordinates on the emergent surface.

In the near-coincidence limit $X^{\prime}\rightarrow X$, Synge\textquoteright s
world function admits a covariant expansion in Riemann normal coordinates.
At leading order, the resulting action reduces to a Polyakov-type
worldsheet action for the separation field $\hat{X}^{a}$, together
with higher-order curvature corrections. Requiring quantum Weyl invariance
produces the leading string beta-function equations for the massless
closed-string fields: the spacetime metric $g_{\mathrm{ab}}$, the
antisymmetric Kalb--Ramond field $B_{\mathrm{ab}}$, and the dilaton
$\phi$.

Because the construction originates from entanglement entropy in AdS$_{3}$,
the resulting background equations must admit an AdS$_{3}$ solution.
In the leading NS--NS equations and for a constant dilaton, a pure
metric background would instead require $R_{\mathrm{ab}}=0$. The
negative curvature of AdS$_{3}$ must therefore be supported by a
nonvanishing three-form flux \cite{Horowitz:1993jc}. This observation
raises a natural question. If $g_{ab}$, $B_{ab}$, and $\phi$ all
emerge from the same entanglement based worldsheet construction, what
are the entanglement counterparts of the Kalb--Ramond field and its
conserved charge?

The answer proposed in our previous work is provided by the bit-thread
formulation of holographic entanglement entropy \cite{Freedman:2016zud,Headrick:2017ucz,Harper:2018sdd}.
On a static bulk slice, a bit-thread flow is a vector field $v^{i}$
satisfying

\begin{equation}
\nabla_{i}v^{i}=0,\qquad|v|\leq\frac{1}{4G_{N}^{\left(3\right)}}.
\end{equation}

\noindent The holographic entanglement entropy can equivalently be
written as a maximum entropy flux through $A$,

\begin{equation}
\Phi\left[A\right]\equiv\int_{A}v^{i}n_{i}dA=\frac{\mathrm{Area}\left(\gamma_{A}\right)}{4G_{N}^{(3)}}=S\left(A\right).
\end{equation}

\noindent On the string side, the Kalb--Ramond field couples to the
string worldsheet \cite{Dabholkar:1990yf,Sen:1992yt,Duff:1994an,Lu:2025awh}.
Varying this coupling with respect to $B_{\mathrm{ab}}$ gives the
antisymmetric spacetime string current

\begin{equation}
j^{\mathrm{ab}}=\frac{1}{\sqrt{-g}}\int d^{2}\sigma\left(\frac{\partial X^{\mathrm{a}}}{\partial\tau}\frac{\partial X^{\mathrm{b}}}{\partial\sigma}-\frac{\partial X^{\mathrm{b}}}{\partial\tau}\frac{\partial X^{\mathrm{a}}}{\partial\sigma}\right)\delta^{\left(3\right)}\left(x-X\left(\tau,\sigma\right)\right).
\end{equation}

\noindent We are interested in the components with one temporal index,
$j^{0i}$, which define a spatial string charge density. This vector
is tangent to the string. Its projection onto a constant-time slice
is

\begin{equation}
q^{i}=N_{\mathrm{ADM}}j^{0i},
\end{equation}

\noindent where $N_{\mathrm{ADM}}$ is the ADM lapse, and we refer
to $q^{i}$ as the projected string charge density. The corresponding
current conservation implies

\begin{equation}
\partial_{i}\left(\sqrt{h}q^{i}\right)=0.
\end{equation}

\noindent After coarse-graining a suitable family of oriented strings,
the projected string charge density can therefore be identified with
bit threads,

\begin{equation}
v^{i}=C_{{\rm geom}}q^{i},
\end{equation}

\noindent where the geometric conversion coefficient $C_{{\rm geom}}$
is fixed by requiring the flow to saturate the standard bit-thread
norm bound $|v|\leq1/\left(4G_{N}^{\left(3\right)}\right)$ at the
bottleneck. In this correspondence, the metric determines the minimal
surface that forms the bottleneck of the flow, whereas the antisymmetric
string current supplies the conserved flux lines passing through it.
Schematically,

\begin{equation}
g_{ab}\quad\longleftrightarrow\quad\text{Ryu-Takayanagi geometry},
\end{equation}

\noindent and

\begin{equation}
B_{ab}\quad\longleftrightarrow\quad\text{Kalb--Ramond charge density}\quad\longleftrightarrow\quad\text{bit threads flow}.
\end{equation}

\begin{figure}[h]
\begin{centering}
\includegraphics[scale=0.35]{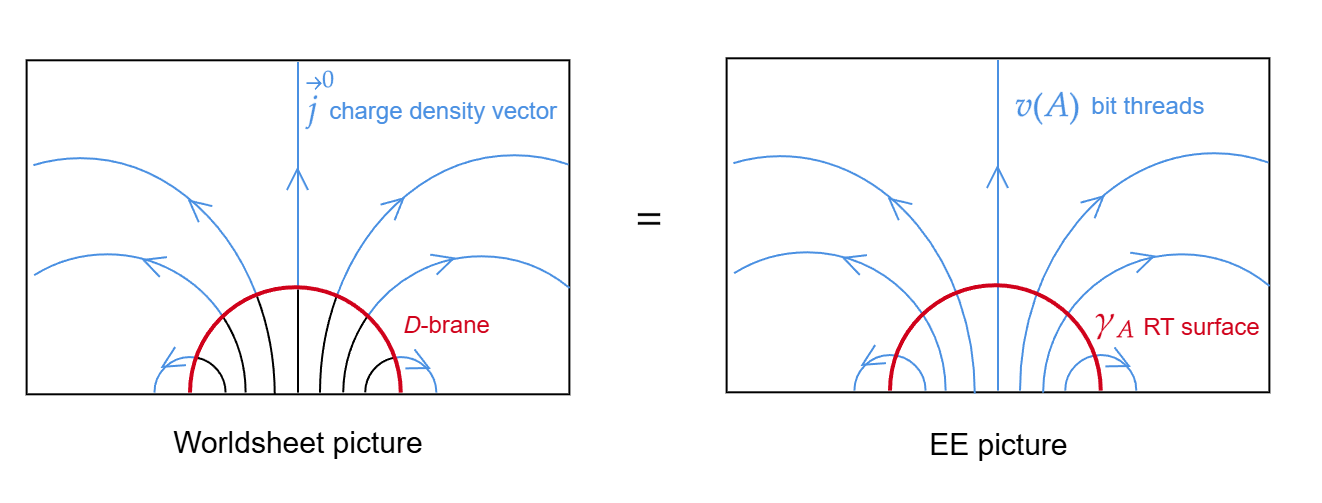}
\par\end{centering}
\caption{\label{fig:bit}This figure shows the connection between the charge
density on the string worldsheet and the bit threads used for entanglement
entropy. In the proposed construction, the entangling brane on which
the open string endpoints are fixed is identified geometrically with
the RT surface $\gamma_{A}$ on a time slice of AdS$_{3}$; this is
a specific auxiliary boundary condition, not a claim that a generic
physical D-brane is an RT surface. To get the AdS$_{3}$ background
from string theory, we need the Kalb-Ramond field. This field gives
a charge density vector $\protect\overrightarrow{j}^{0}$. This vector
points along the open string (which runs from $\sigma=0$ to $\pi$)
and is perpendicular to the D-brane. In the same way, the bit thread
vector $v\left(A\right)$ is perpendicular to the RT surface $\gamma_{A}$.
Both vectors have zero divergence: $\nabla\cdot\protect\overrightarrow{j}^{0}=\nabla\cdot v=0$.
This fact strengthens the correspondence between the worldsheet description
and the bit thread picture of entanglement entropy.}
\end{figure}

\noindent In other words, the coarse-grained Kalb--Ramond charge
density vector $\overrightarrow{j}^{0}$ reproduces the divergenceless
flow structure of holographic bit threads. This establishes an explicit
correspondence between the worldsheet Kalb--Ramond current and the
bit-thread description of entanglement entropy, as illustrated in
figure (\ref{fig:bit}).

In this construction, open string charge reproduces entanglement entropy,
while the corresponding closed string charge reproduces the Bekenstein--Hawking
entropy of the BTZ black hole through open--closed string duality
\cite{Wu:2025qwc}. The locus on which the open-string endpoints are
constrained plays the role of an entangling brane, or E-brane, and
is identified geometrically with the relevant RT surface \cite{Donnelly:2016jet}.
This provides a concrete realization of the relation among the Susskind--Uglum
picture, string charge, and holographic entanglement entropy. Moreover,
although boundary entanglement data can be used to reconstruct an
AdS$_{3}$ bulk geometry and an effective two-dimensional Polyakov-like
action, this reconstruction alone does not select string theory. If
we further interpret this two-dimensional theory as a fundamental
string worldsheet theory, quantum Weyl invariance requires its beta
functions to vanish. For a constant dilaton in the NS--NS sector,
this condition requires a nonzero H-flux to support the AdS$_{3}$
background. Therefore, our construction provides a stringy description
of the entanglement flow, but it does not prove that string theory
is the only possible bulk theory associated with the CFT entanglement
data. Other related work on bit threads can be found in Refs. \cite{Agon:2018lwq,Caggioli:2024uza,Lin:2026ehl,Caceres:2025ypk}

The above construction was developed primarily in three-dimensional,
locally AdS$_{3}$ backgrounds. The \textbf{first aim} of the present
paper is to determine whether it extends consistently to ten-dimensional
superstring backgrounds and whether it can reproduce the entropy of
a higher-dimensional black hole. Because the correspondence is formulated
in terms of fundamental-string charge, the natural setting is not
an arbitrary higher-dimensional gravitational theory but Type IIB
supergravity with explicit string probes.

A canonical testing ground is the fundamental-string (F1)--Neveu--Schwarz
five-brane (NS5)--momentum (P) system \cite{Tseytlin:1996bh,Mathur:2005zp,Martinec:2017ztd,Martinec:2018nco}.
In ten dimensions, this system consists of fundamental strings and
NS5-branes sharing a common direction $w$, together with momentum
$P$ propagating along that direction. The NS5-branes additionally
wrap an internal $T^{4}$. When $w$ and $T^{4}$ are compactified,
the configuration gives a five-dimensional three-charge black hole. 

In the F1--NS5--P duality frame, the relevant ten-dimensional string-frame
action can be written as

\begin{equation}
S=\frac{1}{2\kappa_{10}^{2}}\int d^{10}x\sqrt{-g}e^{-2\phi}\left(R+4\left(\nabla\phi\right)^{2}-\frac{1}{12}\left(H^{\left(3\right)}\right)^{2}\right)+S_{\mathrm{F1}},
\end{equation}

\noindent where the additional term $S_{\mathrm{F1}}$ describes radial
F1 probes, which should be distinguished from the background foundamental
string charge already encoded in the supergravity flux. At leading
order in the probe approximation, the background is given by the extremal
F1--NS5--P solution

\begin{eqnarray}
dS_{\mathrm{F1-NS5-P}}^{2} & = & \frac{1}{H_{1}}\left[-dt^{2}+dw^{2}+K\left(dt+dw\right)^{2}\right]+H_{5}\left(dr^{2}+r^{2}d\Omega_{3}^{2}\right)+\delta_{ij}dy^{i}dy^{j},\nonumber \\
e^{2\phi\left(r\right)} & = & \frac{H_{5}}{H_{1}},\qquad H=d\left(1-H_{1}^{-1}\right)\wedge dt\wedge dw+2Q_{5}\omega_{3},
\end{eqnarray}

\noindent with

\begin{equation}
H_{1}\left(r\right)=1+\frac{Q_{1}}{r^{2}},\qquad H_{5}\left(r\right)=1+\frac{Q_{5}}{r^{2}},\qquad K\left(r\right)=\frac{Q_{P}}{r^{2}}.
\end{equation}

\noindent Here $\omega_{3}$ is the volume form of the unit three-sphere.
The electric component of the Kalb-Ramond field strength $H$ carries
the F1 charge, the magnetic component carries the NS5 charge, and
$K\left(r\right)$ describes the chiral momentum wave. In the near-horizon
limit, the geometry becomes locally \cite{Martinec:2017ztd}

\begin{equation}
\mathrm{BTZ}_{3}\times S^{3}\times T^{4},
\end{equation}

\noindent when the momentum charge is nonzero and the common direction
$w$ is compact. This structure provides a direct connection between
the ten-dimensional system and the BTZ construction developed previously.

We treat the additional radial F1 strings in the probe approximation.
This means that the dimensionless combination controlling their backreaction
is taken to be small compared with the scales set by the background
charges. We thus solve the source-free supergravity equations in the
exterior region and then evaluate the conserved probe current on that
fixed background. A fully backreacted configuration would require
solving the coupled supergravity and source equations and is beyond
the scope of the present analysis.

Next, we generalize the previous $\mathcal{N}$ parallel string probes
distributed uniformly along a one-dimensional RT surface to the higher-dimensional
case: we place $\mathcal{N}$ radially extended F1 string probes uniformly
over the compact eight-dimensional (co-dimension two) horizon section
$\Sigma_{8}=S_{w}^{1}\times S^{3}\times T^{4}$. Each radial string
pierces one of $\mathcal{N}$ equal cells of the horizon section.
In this configuration, we calculate the F1 charge-density current,
whose only nonvanishing component is $j_{\mathrm{F1}}^{0r}$. Its
projection onto a constant-time slice is

\begin{equation}
q_{\mathrm{F1}}^{I}=N_{\mathrm{ADM}}j_{\mathrm{F1}}^{0I},\qquad\partial_{I}\left(\sqrt{h}q_{\mathrm{F1}}^{I}\right)=0,
\end{equation}

\noindent where $I$ labels the spatial directions. To relate this
current to bit threads, we construct the radial flow

\begin{equation}
v_{\mathrm{F1}}^{I}=C_{{\rm geom}}^{\mathrm{F1}}q_{\mathrm{F1}}^{I},
\end{equation}

\noindent where $C_{{\rm geom}}^{\mathrm{F1}}$ can be fixed by the
corresponding string-frame generalization of the bit-thread norm bound:

\begin{equation}
\left|v_{\mathrm{F1}}\left(r\right)\right|\leq\frac{e^{-2\phi\left(r\right)}}{4G_{N}^{\left(10\right)}}.
\end{equation}

\noindent The dilaton factor arises because the bit-thread norm bound
is naturally defined in the Einstein frame and acquires the factor
$\exp\left(-2\phi\right)$ when expressed in the string frame. Requiring
the radial flow to saturate this bound at the minimal horizon section
fixes $C_{{\rm geom}}^{\mathrm{F1}}$. Its entropy flux then gives

\begin{equation}
\Phi_{\mathrm{F1}}=\int v_{\mathrm{F1}}^{r}n_{r}dS=\frac{L_{w}V_{T^{4}}2\pi^{2}}{4G_{N}^{\left(10\right)}}\sqrt{Q_{1}Q_{5}Q_{P}}=S_{\mathrm{BH}}^{\mathrm{F1}},
\end{equation}

\noindent which agrees exactly with the standard Bekenstein--Hawking
entropy \cite{Cardy:1986ie,Strominger:1996sh,Callan:1996dv}.

A further advantage of this framework is that it allows us to study
how the charge density description of bit threads transforms under
string dualities. As we showed in our previous work, if the correspondence
between worldsheet string charge density and entanglement bit threads
is fundamental, dualities of string theory should also have corresponding
descriptions in terms of entanglement observables. For example, open--closed
string duality is related, in our construction, to the Susskind--Uglum
interpretation of black-hole entropy. In Type IIB supergravity, the
F1--NS5--P and D1--D5--P \cite{Cvetic:1995uj,Tseytlin:1996as,Horowitz:1996ay}
systems are related by S-duality \cite{Mathur:2005zp}:

\begin{equation}
\mathrm{F1}\longleftrightarrow\mathrm{D1},\qquad\mathrm{NS5}\longleftrightarrow\mathrm{D5},\qquad\mathrm{P}\longleftrightarrow\mathrm{P}.
\end{equation}

\noindent The fundamental string F1, which couples electrically to
$B$, is mapped to an object that couples electrically to $C^{\left(2\right)\prime}$,
namely the D1-brane. The NS5-brane is the magnetic source for $B$.
After S-duality, this magnetic NS--NS flux becomes R--R flux, whose
magnetic source is the D5-brane. The momentum charge is carried by
the metric and is neutral under S-duality. The microscopic state counting
of the D1--D5--P system provides one of the classic successes of
string theory \cite{Cardy:1986ie,Strominger:1996sh,Callan:1996dv,Giusto:2010gv}.

Therefore, the \textbf{second aim} of this paper is to determine how
this construction behaves under the exact S-duality of Type IIB string
theory. Because the Bekenstein--Hawking entropy is a physical quantity,
the F1--NS5--P and D1--D5--P descriptions must give the same result
when all fields and compactification data are transformed consistently.
This result provides a nontrivial consistency check of the proposed
string charge density/bit threads correspondence. Under S-duality,
the Kalb--Ramond current of $S_{\mathrm{F1}}$ is mapped to an R--R
D1 current of $S_{\mathrm{D1}}$,

\begin{equation}
j_{\mathrm{F1}}^{\mathrm{ab}}\quad\xrightarrow{\quad S\quad}\quad j_{\mathrm{D1}}^{\mathrm{ab}}.
\end{equation}

\noindent The resulting D1 charge density again defines a divergenceless
spatial flow,

\begin{equation}
v_{\mathrm{D1}}^{I}=C_{{\rm geom}}^{\mathrm{D1}}q_{\mathrm{D1}}^{I}.
\end{equation}

\noindent The local coordinate components of $v_{\mathrm{F1}}^{I}$
and $v_{\mathrm{D1}}^{I}$ are not equal because the string-frame
metric, dilaton, spatial measure, and coordinate current components
transform under S-duality. For matched integer source normalization,
the geometric conversion coefficients $C_{{\rm geom}}^{\mathrm{F1}}$
and $C_{{\rm geom}}^{\mathrm{D1}}$ in the F1 and D1 frames are equal,
as shown explicitly in Section 7. Therefore, the physical invariant
is the integrated entropy flux (black hole entropy) rather than an
individual coordinate component. Repeating the above calculation in
the D1--D5--P frame, we show that the two duality frames produce
the same entropy, as expected:

\begin{equation}
\Phi_{\mathrm{F1}}=\Phi_{\mathrm{D1}}=S_{\mathrm{BH}}=\frac{L_{w}V_{T^{4}}2\pi^{2}}{4G_{N}^{\left(10\right)}}\sqrt{Q_{1}Q_{5}Q_{P}}.
\end{equation}

\noindent Thus, S-duality does not leave the local string-frame representation
of the bit threads unchanged. Rather, it maps an NS--NS realization
of the entropy flow into an R--R realization while preserving the
maximal physical flux.

On the other hand, as discussed above, the geometric conversion coefficient
$C_{{\rm geom}}$ can be fixed by the standard bit-thread norm bound
in the bulk. Because of the correspondence between the string charge
density and bit threads, the same quantities may also be obtained
from string worldsheet or the corresponding CFT. The D1--D5 CFT lives
at the asymptotic conformal boundary of the decoupled AdS$_{3}$ throat,
and the entanglement entropy of the boundary region is known from
the Cardy formula. Therefore, the \textbf{third aim} of this paper
is to ask whether the microscopic description of the CFT can independently
determine the CFT conversion coefficient $C_{{\rm CFT}}$, and to
clarify the physical meanings of $C_{{\rm geom}}$ and $C_{{\rm CFT}}$. 

Specifically, the D1--D5 CFT lives at the asymptotic conformal boundary
of the decoupled $\mathrm{AdS}_{3}$ throat. For one complete boundary
factor in a doubled microcanonical state, the entanglement entropy
is fixed by the Cardy degeneracy. Taking a microcanonical purification
of the black hole sector \cite{Azeyanagi:2007bj}, 

\begin{equation}
|\Psi_{{\rm micro}}\rangle=\frac{1}{\sqrt{\Omega}}\sum_{a=1}^{\Omega}|a\rangle_{L}|a\rangle_{R},\qquad S\left(R\right)=\log\Omega,
\end{equation}

\noindent one obtains 

\begin{equation}
S_{{\rm CFT}}\left(R\right)=2\pi\sqrt{\frac{c_{L}N_{L}}{6}}=2\pi\sqrt{n_{1}n_{5}n_{P}},
\end{equation}

\noindent where $R$ is the entire right CFT, $c_{L}=6n_{1}n_{5}$,
and $N_{L}=n_{P}$ in the Cardy regime \cite{Cardy:1986ie,Strominger:1996sh,Callan:1996dv}. 

Motivated by the string charge density/bit threads correspondence,
we formulate the D1--D5 \emph{boundary-matching conjecture},

\begin{equation}
\Phi_{{\rm D1}}\left(R\right)\equiv C_{{\rm CFT}}\mathcal{N}=S_{{\rm CFT}}\left(R\right).\label{eq:CFTconjecture}
\end{equation}

\noindent This is the only conjectural input in the microscopic normalization.
It states that the boundary state and the complete boundary subsystem
fix the total radial entropy flux. It does not identify the geometrical
size of a generic boundary interval with a number of strings. Since
the D1 current has no endpoints between the throat boundary and the
horizon, current conservation makes the same integrated flux cross
every homologous radial section.

Using the charge dictionary, the conjecture gives
\begin{equation}
C_{{\rm CFT}}=\frac{\mathcal{V}_{0}}{4G_{N}^{\left(10\right)}\mathcal{N}}\sqrt{Q_{1}Q_{5}Q_{P}}=C_{{\rm geom}}^{\mathrm{F1/D1}}.\label{eq:microequalsarea}
\end{equation}
Thus, $C_{{\rm CFT}}$ reproduces the geometric coefficient rather
than assuming it. Moreover, the CFT-normalized D1 flow satisfies
\begin{align}
\frac{|v_{\mathrm{D1}}\left(r\right)|}{e^{-2\phi^{\prime}\left(r\right)}/\left(4G_{N}^{\left(10\right)}\right)} & =\left[\frac{Q_{1}Q_{5}Q_{P}}{\prod_{a=1,5,P}(r^{2}+Q_{a})}\right]^{1/2}\nonumber \\[-2pt]
 & \leq1.\label{eq:derivedbound}
\end{align}
For positive charges, equality is approached only in the exterior
extremal-horizon limit $r\rightarrow0^{+}$. Therefore, conditional
on the \emph{boundary-matching conjecture}, the CFT normalization
reproduces the expected string-frame norm bound throughout the homogeneous
radial sector and fixes the same bottleneck as the geometric construction.

This comparison also gives a simple physical interpretation of the
conversion coefficients:

\begin{equation}
C_{{\rm CFT}}=\frac{S_{{\rm CFT}}\left(R\right)}{\mathcal{N}}=\frac{S_{{\rm BH}}}{\mathcal{N}}=C_{{\rm geom}}.
\end{equation}

\noindent For the homogeneous coarse-grained current considered here,
$C_{{\rm geom}}$ and $C_{{\rm CFT}}$ measure the entropy-flux capacity
associated with one unit of the normalized string-charge congruence.
The same capacity can therefore be fixed from either the boundary
entanglement entropy or the bulk horizon. Equivalently, once this
capacity is fixed, the total entropy determines how many string-charge
tubes are required to carry the maximal flux. This provides an alternative
way to understand the relation between boundary entanglement entropy
and black hole entropy, and suggests a possible connection between
string-charge transport and the encoding of information in holographic
systems.

These results show that the information encoded by the maximal flow
is duality invariant even though its auxiliary source-current representation
depends on the chosen duality frame. They further suggest a possible
extension to the full $SL\left(2,\mathbb{Z}\right)$ family of $\left(p,q\right)$-string
charge flows \cite{Schwarz:1995dk,Townsend:1997kr,Bergshoeff:2006gs}.

\begin{figure}[h]
\begin{centering}
\includegraphics[scale=0.35]{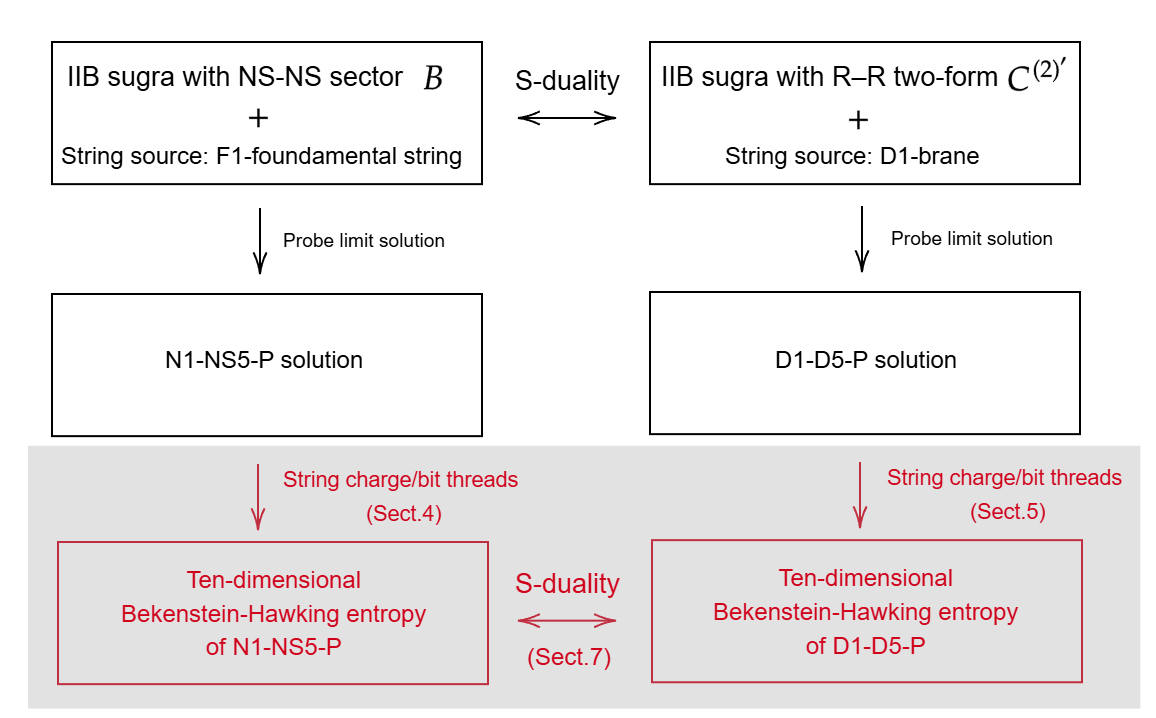}
\par\end{centering}
\caption{\label{fig:aim}The unshaded region denotes known results from previous
works. In this paper, we aim to calculate the Bekenstein--Hawking
entropies of the F1--NS5--P and D1--D5--P solutions, and study
their S-duality, as shown in the shaded region.}
\end{figure}

The main results of this paper are therefore to generalize the string
charge density/bit thread correspondence to ten dimensions, calculate
the Bekenstein--Hawking entropies of the F1--NS5--P and D1--D5--P
systems, provide a conjecture-based CFT determination of an independent
normalization and compare it with the geometric normalization, test
the resulting flow pointwise against the expected norm bound, and
study the transformation of the charge flows under S-duality. These
aims and results are illustrated in figure (\ref{fig:aim}).

Finally, it is worth noting that the analysis is performed in the
probe and continuum-smearing approximations. A fully backreacted treatment
would require solving the coupled supergravity and source equations,
while a manifestly $SL\left(2,\mathbb{Z}\right)$ covariant extension
would require the axion and the complete Type IIB field content. Subject
to these limitations, the agreement between the F1 and D1 descriptions
provides a nontrivial consistency check that the proposed string charge
density/bit threads correspondence is compatible with the fundamental
nonperturbative S-duality of Type IIB string theory.

The remainder of the paper is organized as follows. In Section 2,
we review the string charge density/bit threads correspondence and
demonstrate its application to the Bekenstein--Hawking entropy of
the BTZ black hole. In Section 3, we review the extremal F1--NS5--P
and D1--D5--P solutions, their near-horizon geometries, and their
relation under S-duality. In Section 4, we construct the smeared radial
F1 current and calculate the F1--NS5--P entropy. In Section 5, we
perform the corresponding analysis for the S-dual D1--D5--P system.
In Section 6, we set aside the geometric normalization and use the
D1--D5 CFT together with an explicit boundary-matching conjecture
to determine $C_{{\rm CFT}}$ independently, compare it with $C_{{\rm geom}}$,
and test the resulting norm bound pointwise throughout the exterior
geometry. In Section 7, we analyze the transformation of the charge
density flows under S-duality and identify the duality invariant entropy
flux. The final section summarizes our conclusions and discusses directions
for future work.

\section{String charge density/bit threads correspondence}

To illustrate how the Kalb--Ramond charge density/bit threads correspondence
can be used to calculate the Bekenstein--Hawking entropy of a BTZ
black hole, we first consider an example that was not explicitly discussed
in our previous work. Here we do not invoke open--closed string duality;
instead, we calculate the BTZ entropy directly from the Kalb--Ramond
charge density and its corresponding bit-thread flow. We begin with
a three-dimensional low energy effective action for the NS--NS sector
of closed string theory, supplemented by a macroscopic fundamental-string
source,

\begin{equation}
S=\frac{1}{2\kappa_{3}^{2}}\int d^{3}x\sqrt{-g}e^{-2\phi}\left(R+4\left(\nabla\phi\right)^{2}-\frac{1}{12}\left(H^{\left(3\right)}\right)^{2}+\frac{4}{l_{AdS}^{2}}\right)+S_{\sigma},
\end{equation}

\noindent where $H^{\left(3\right)}=dB^{\left(2\right)}$, $\kappa_{3}^{2}=8\pi G_{N}^{\left(3\right)}$,
and the string source is described by the sigma-model action

\begin{equation}
S_{\sigma}=-\frac{1}{4\pi\alpha^{\prime}}\int d^{2}\sigma\left(\sqrt{-\gamma}\gamma^{mn}\partial_{m}X^{\mathrm{a}}\partial_{n}X^{\mathrm{b}}g_{\mathrm{ab}}+\epsilon^{mn}\partial_{m}X^{\mathrm{a}}\partial_{n}X^{\mathrm{b}}B_{\mathrm{ab}}\right),
\end{equation}

\noindent This action describes a macroscopic string carrying Kalb--Ramond
charge and propagating in a dynamical background consisting of the
metric, dilaton, and Kalb--Ramond field. The bulk equations of motion
are obtained by varying the action with respect to these fields:

\begin{eqnarray}
R^{\mathrm{ab}}+2\nabla^{\mathrm{a}}\nabla^{\mathrm{b}}\phi-\frac{1}{4}H^{\mathrm{acd}}H_{\:\mathrm{cd}}^{\mathrm{b}} & = & \kappa_{3}^{2}T^{\mathrm{ab}},\nonumber \\
\nabla_{\mathrm{a}}\left(e^{-2\phi}H^{\mathrm{abc}}\right) & = & \frac{\kappa_{3}^{2}}{\pi\alpha^{\prime}}j^{\mathrm{bc}},\nonumber \\
4\nabla^{2}\phi-4\left(\nabla\phi\right)^{2}+R-\frac{1}{12}H^{2}+\frac{4}{l_{AdS}^{2}} & = & 0.
\end{eqnarray}

\noindent The source terms entering these equations are

\begin{eqnarray}
T^{\mathrm{ab}} & = & -\frac{e^{2\phi}}{2\pi\alpha^{\prime}\sqrt{-g}}\int d^{2}\sigma\sqrt{-\gamma}\gamma^{mn}\partial_{m}X^{\mathrm{a}}\partial_{n}X^{\mathrm{b}}\delta^{\left(3\right)}\left(x-X\left(\tau,\sigma\right)\right),\nonumber \\
j^{\mathrm{ab}} & = & \frac{1}{\sqrt{-g}}\int d^{2}\sigma\left(\frac{\partial X^{\mathrm{a}}}{\partial\tau}\frac{\partial X^{\mathrm{b}}}{\partial\sigma}-\frac{\partial X^{\mathrm{b}}}{\partial\tau}\frac{\partial X^{\mathrm{a}}}{\partial\sigma}\right)\delta^{\left(3\right)}\left(x-X\left(\tau,\sigma\right)\right).\label{eq: source}
\end{eqnarray}

To determine the background geometry, we work in the probe approximation
\cite{Gubser:2002tv,Chang:2013mca}. The radial probe string has codimension
one in three dimensions. Locally denoting the proper transverse coordinate
by $x_{\bot}$, the perturbation $u$ of the metric, dilaton, or two-form
sourced by string satisfies schematically

\begin{equation}
\partial_{\bot}^{2}u\left(x_{\bot}\right)\sim\kappa_{3}^{2}T_{\mathrm{probe}}\delta\left(x_{\bot}\right),\qquad T_{\mathrm{probe}}=1/\left(2\pi\alpha^{\prime}\right).
\end{equation}

\noindent The Green function of the one-dimensional Laplacian obeys

\begin{equation}
\frac{d^{2}}{dx_{\bot}^{2}}\left|x_{\bot}\right|=2\delta\left(x_{\bot}\right).
\end{equation}

\noindent The probe-induced perturbation therefore grows linearly
with the proper transverse distance,

\begin{equation}
u\left(x_{\bot}\right)\sim\kappa_{3}^{2}T_{\mathrm{probe}}\left|x_{\bot}\right|\sim G_{N}^{(3)}T_{\mathrm{probe}}\left|x_{\bot}\right|,
\end{equation}

\noindent up to numerical coefficients. Evaluated at a characteristic
background scale $L$, the local backreaction is parametrically controlled
by

\begin{equation}
\epsilon_{\mathrm{probe}}^{\left(3\right)}\sim G_{N}^{(3)}T_{\mathrm{probe}}L.
\end{equation}

\noindent The probe approximation consequently requires

\begin{equation}
\epsilon_{\mathrm{probe}}^{\left(3\right)}\ll1.
\end{equation}

\noindent These limits should be understood as suppressing the backreaction
terms in the bulk equations of motion. The worldsheet current itself
is retained as a kinematical probe quantity and will subsequently
be used to construct a flow on the fixed background. Thus, one first
solves the source-free background equations and then evaluates the
probe string current on that solution.

This procedure is closely analogous, although not identical, to the
cosmic brane construction of holographic Renyi entropy developed by
Xi Dong \cite{Dong:2016fnf}, building on the gravitational replica
method \cite{Lewkowycz:2013nqa}. In that construction, the cosmic
brane has tension $\frac{n-1}{4nG_{N}}$, and therefore becomes tensionless
in the entanglement entropy limit $n\rightarrow1$. Its backreaction
then vanishes, and its worldvolume approaches the Ryu--Takayanagi
minimal surface. This provides a useful precedent for extracting geometric
information from a source whose backreaction is negligible in the
relevant limit.

In the present case, the source-free equations admit the nonrotating
BTZ solution supported by NS--NS flux \cite{Horowitz:1993jc}

\noindent 
\begin{eqnarray}
dS_{\mathrm{BTZ}}^{2} & = & -\frac{r_{+}^{2}}{l_{AdS}^{2}}\sinh^{2}\rho dt^{2}+l_{AdS}^{2}d\rho^{2}+r_{+}^{2}\cosh^{2}\rho d\theta^{2},\nonumber \\
H_{\mathrm{abc}} & = & \frac{2}{l_{AdS}}\epsilon_{\mathrm{abc}},\qquad B_{t\theta}=-\frac{r_{+}^{2}}{l_{AdS}}\sinh^{2}\rho,\nonumber \\
\phi & = & \phi_{0}=\mathrm{const}.
\end{eqnarray}

\noindent The coordinate $\theta$ has periodicity $\theta\sim\theta+2\pi$,
and the minimal circle at $\rho=0$, with circumference $2\pi r_{+}$,
is the event horizon of the black hole. The time coordinate has been
normalized so that this metric describes the general nonrotating BTZ
geometry with arbitrary $r_{+}$. Moreover, this metric describes
one exterior region of the nonrotating BTZ black hole when $\rho\geq0$.
The second exterior region is represented by an identical copy. At
$t=0$, the two copies can be combined by introducing a signed proper-distance
coordinate $\rho\in\left(-\infty,+\infty\right)$, with $\rho>0$
and $\rho<0$ corresponding to the right and left exteriors, respectively
\cite{Lin:2025jjh}. The bifurcation circle is located at $\rho=0$.
The complete Lorentzian two-sided geometry, including the interior
regions, requires Kruskal coordinates. Consider a radially oriented
macroscopic string in the static gauge

\begin{equation}
X^{t}=\tau,\qquad X^{\rho}=\sigma,\qquad X^{\theta}=\theta_{o},
\end{equation}

\noindent Substituting this embedding into the definition of the antisymmetric
current (\ref{eq: source}) gives

\begin{equation}
j_{\mathrm{BTZ}}^{0\rho}=\frac{1}{\sqrt{-g}}\delta_{2\pi}\left(\theta-\theta_{o}\right),
\end{equation}

\noindent where $\delta_{2\pi}$ denotes the periodic delta function
on the $\theta$-circle. We now consider $\mathcal{N}$ parallel macroscopic
strings distributed uniformly along the $\theta$-direction. The corresponding
$\mathcal{N}$-string probe limit is

\begin{equation}
\epsilon_{\mathrm{probe}}^{\left(3\right)}\sim G_{N}^{\left(3\right)}T_{\mathrm{probe}}\mathcal{N}L\ll1.
\end{equation}

\noindent Accordingly, the continuum description means coarse-graining
at large but finite $\mathcal{N}$ within the window $1\ll\mathcal{N}\ll\left[G_{N}^{\left(3\right)}T_{\mathrm{probe}}L\right]^{-1}$;
a literal $\mathcal{N}\rightarrow\infty$ limit at fixed tension would
leave the probe regime. Multi-string solutions and configurations
of parallel macroscopic strings have been extensively discussed in
the literature \cite{Dabholkar:1990yf,Sen:1992yt,Duff:1994an}. The
corresponding current is

\begin{equation}
j^{0\rho}=\frac{1}{\sqrt{-g}}\stackrel[k=0]{\mathcal{N}-1}{\sum}\delta_{2\pi}\left(\theta-\frac{2\pi k}{\mathcal{N}}\right),
\end{equation}

\noindent The angular separation between adjacent strings is $\theta_{s}=2\pi/\mathcal{N}$,
so that

\begin{equation}
j_{\mathrm{BTZ}}^{0\rho}=\frac{1}{\sqrt{-g}}\frac{1}{\theta_{s}}\left[\theta_{s}\stackrel[k=0]{\mathcal{N}-1}{\sum}\delta_{2\pi}\left(\theta-k\theta_{s}\right)\right]=\frac{1}{\sqrt{-g}}\frac{\mathcal{N}}{2\pi}\left[\theta_{s}\stackrel[k=0]{\mathcal{N}-1}{\sum}\delta_{2\pi}\left(\theta-k\theta_{s}\right)\right].
\end{equation}

\begin{figure}[h]
\begin{centering}
\includegraphics[scale=0.15]{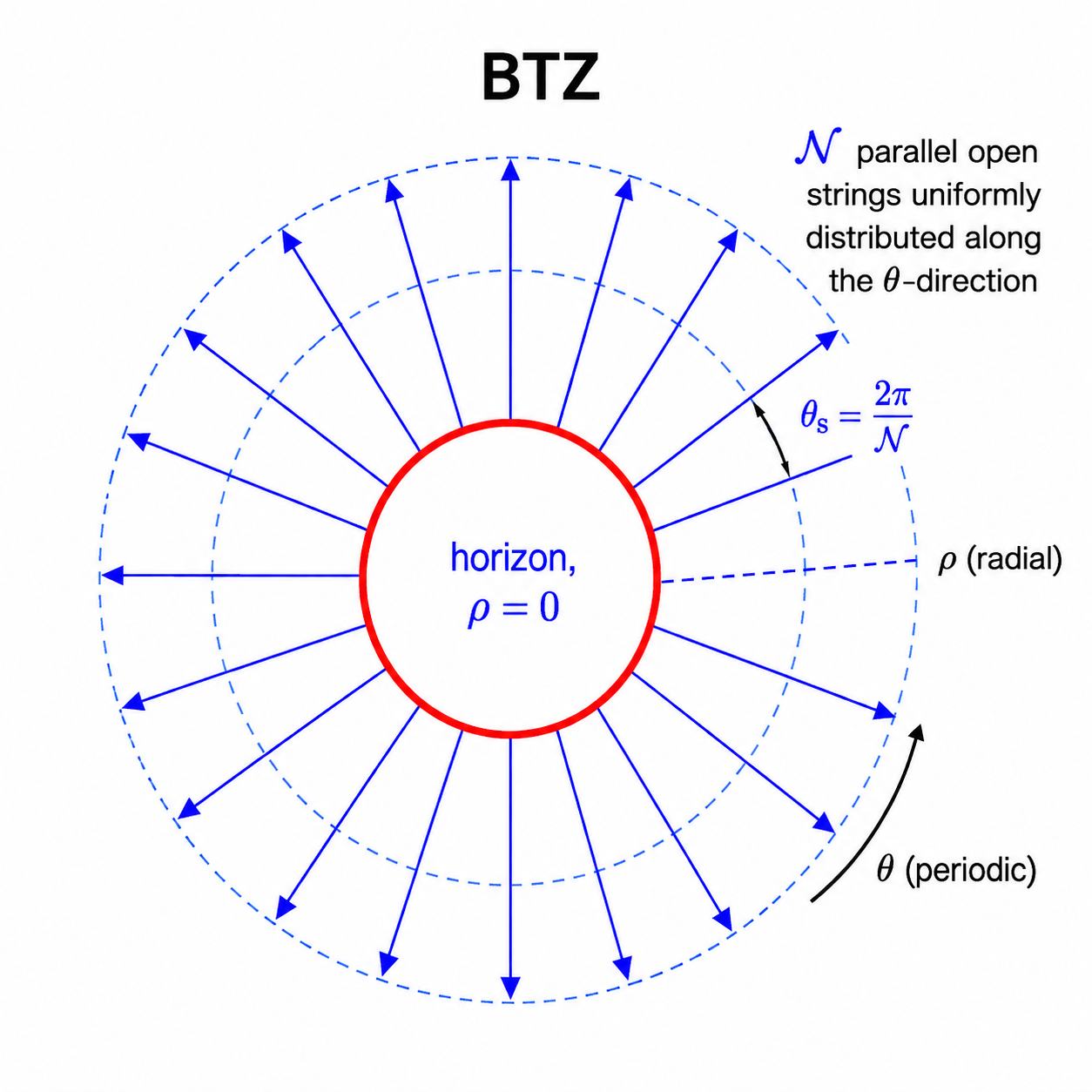}
\par\end{centering}
\caption{\label{fig:BTZ}This figure illustrates how $\mathcal{N}$ parallel
macroscopic strings are uniformly distributed along the angular direction
$\theta$ on one side of the BTZ black hole. The red circle represents
the horizon, located at $\rho=0$, with circumference $2\pi r_{+}$.
The blue lines with arrows denote the string charge density, which
emanates from the horizon toward the asymptotic boundary. The angular
separation between adjacent strings is $\theta_{s}=2\pi/\mathcal{N}$.
This implies that each string crosses a one-dimensional angular cell
of size $\theta_{s}$.}
\end{figure}

\noindent The expression in square brackets tends to unity in the
continuum limit, while $\mathcal{N}/\left(2\pi\right)$ is the coarse-grained
angular line density. The configuration of $\mathcal{N}$ parallel
macroscopic strings is illustrated in figure (\ref{fig:BTZ}). 
In the formal distributional continuum limit, $\mathcal{N}\rightarrow\infty$
(hence $\theta_{s}\rightarrow0$), the quantity in square brackets
converges to unity in the distributional sense:

\begin{equation}
\theta_{s}\sum_{k=0}^{\mathcal{N}-1}\delta_{2\pi}\left(\theta-k\theta_{s}\right)\longrightarrow1.
\end{equation}

\noindent Consequently, the total string charge density approaches

\begin{equation}
j_{\mathrm{BTZ}}^{0\rho}=\frac{1}{\sqrt{-g}}\frac{\mathcal{N}}{2\pi}=\frac{2}{r_{+}^{2}\sinh\left(2\rho\right)}\frac{\mathcal{N}}{2\pi}.
\end{equation}

\noindent Returning to the BTZ metric, the resulting string charge
density vector can thus be written as

\begin{equation}
j_{\mathrm{BTZ}}^{\mathrm{0a}}=\left(j_{\mathrm{BTZ}}^{0t},j_{\mathrm{BTZ}}^{0\rho},j_{\mathrm{BTZ}}^{0\theta}\right)=\frac{2}{r_{+}^{2}\sinh\left(2\rho\right)}\frac{\mathcal{N}}{2\pi}\left(0,1,0\right),\qquad\mathrm{a}=t,\rho,\theta.
\end{equation}

\noindent As argued in our previous work \cite{Wu:2025qwc}, it is
sufficient to perform the calculation in one exterior region. The
second exterior gives an identical result by the reflection symmetry
of the two-sided BTZ geometry. When two identical arrays of parallel
open-string segments are distributed uniformly over the horizon, their
numbers and intersection points can be matched pairwise across the
throat. Provided that their orientations are chosen consistently,
the corresponding segments can be smoothly joined into worldsheets
extending through the two-sided geometry. Equivalently, because the
outward-pointing normals on the two sides have opposite orientations,
smooth gluing requires

\begin{eqnarray}
\overrightarrow{j}^{0}\left(\rho\right) & = & \overrightarrow{j}_{+}^{0}\left(\rho_{+}\right)\cup\overrightarrow{j}_{-}^{0}\left(\rho_{-}\right)\nonumber \\
 & = & \int d\sigma^{+}\delta\left(\overrightarrow{x}-\overrightarrow{X}_{+}\left(\tau,\sigma^{+}\right)\right)\partial_{\sigma^{+}}\overrightarrow{X}_{+}\left(\tau,\sigma^{+}\right)\nonumber \\
 &  & +\int d\sigma^{-}\delta\left(\overrightarrow{x}-\overrightarrow{X}_{-}\left(\tau,\sigma^{-}\right)\right)\partial_{\sigma^{-}}\overrightarrow{X}_{-}\left(\tau,\sigma^{-}\right)\nonumber \\
 & = & \int d\sigma\delta\left(\overrightarrow{x}-\overrightarrow{X}\left(\tau,\sigma\right)\right)\partial_{\sigma}\overrightarrow{X}\left(\tau,\sigma\right),
\end{eqnarray}

\noindent This matching condition ensures that no additional delta
function source is generated at the gluing surface and that the complete
current satisfies the divergenceless equation distributionally throughout
the two-sided geometry:

\begin{equation}
\partial_{\mathrm{a}}\left(\sqrt{-g}j_{\mathrm{BTZ}}^{\mathrm{0a}}\right)=0.
\end{equation}

\noindent In the remainder of this paper, we therefore focus on the
result in one exterior region; the calculation in the other region
is identical and straightforward.

To compare this current with bit threads, we restrict it to a constant-time
spatial slice $\Sigma$. Writing the spacetime metric in ADM form,
we define the projected string charge density on the time slice by

\begin{equation}
q_{\mathrm{BTZ}}^{i}=N_{\mathrm{ADM}}j_{\mathrm{BTZ}}^{0i}=\frac{\sqrt{-g}}{\sqrt{h}}j_{\mathrm{BTZ}}^{0i}=\sqrt{-g_{tt}}j_{\mathrm{BTZ}}^{0i},\qquad i=\rho,\theta,
\end{equation}

\noindent where $N_{\mathrm{ADM}}=\sqrt{-g_{tt}}$ is the lapse, such
that $\sqrt{-g}=N_{\mathrm{ADM}}\sqrt{h}$. It follows immediately
that

\begin{equation}
\partial_{i}\left(\sqrt{h}q_{\mathrm{BTZ}}^{i}\right)=0.
\end{equation}

\noindent Explicitly, the projected string charge density is

\begin{equation}
q_{\mathrm{BTZ}}^{i}=\sqrt{-g_{tt}}j_{\mathrm{BTZ}}^{0i}=\frac{1}{l_{AdS}r_{+}}\frac{1}{\cosh\rho}\frac{\mathcal{N}}{2\pi}\left(1,0\right).\label{eq:open string charge-1-1}
\end{equation}

\noindent We now identify a candidate bit-thread flow proportional
to this induced Kalb--Ramond charge density:

\begin{equation}
v_{\mathrm{BTZ}}^{i}=C_{{\rm geom}}^{\mathrm{BTZ}}q_{\mathrm{BTZ}}^{i}=\left(v_{\mathrm{BTZ}}^{\rho},v_{\mathrm{BTZ}}^{\theta}\right)=C_{{\rm geom}}^{\mathrm{BTZ}}\frac{1}{l_{AdS}r_{+}}\frac{1}{\cosh\rho}\frac{\mathcal{N}}{2\pi}\left(1,0\right).
\end{equation}

\noindent This vector field is divergenceless by construction. Its
norm with respect to the induced metric is

\begin{equation}
\left|v_{\mathrm{BTZ}}\right|=\sqrt{h_{ij}v_{\mathrm{BTZ}}^{i}v_{\mathrm{BTZ}}^{j}}=C_{{\rm geom}}^{\mathrm{BTZ}}\frac{\mathcal{N}}{2\pi r_{+}}\frac{1}{\cosh\rho}.
\end{equation}

\noindent According to the bit threads formulation of holographic
entanglement entropy, the flow must satisfy the bit-threads norm bound
$\left|v_{\mathrm{BTZ}}\right|\leq1/4G_{N}^{\left(3\right)}$ \cite{Freedman:2016zud,Headrick:2017ucz,Harper:2018sdd}.
In our case, we need to generalize this bound to the string-frame
solution including the dilaton:

\begin{equation}
\left|v_{\mathrm{BTZ}}\right|\leq\frac{e^{-2\phi}}{4G_{N}^{\left(3\right)}},
\end{equation}

\noindent where $\phi=\phi_{0}$ is constant and $\phi_{0}$ can be
set to zero in the asymptotically normalized dilaton convention. In
the bit threads formulation, the RT entropy can be reinterpreted as
the maximum flux of a divergenceless vector field $v^{i}$ through
the boundary entangling region, dual to the minimal bulk surface area.
The bound implements the local capacity constraint: it limits the
density of threads such that at most one independent EPR pair can
pass through an area element of size $4G_{N}^{\left(3\right)}$. Because
$\frac{1}{\cosh\rho}\leq1$, the norm is maximal at the throat $\rho=0$.
Requiring the flow to saturate the max-flow/min-cut bound there gives
$1/4G_{N}^{\left(3\right)}$, and hence 

\begin{equation}
C_{{\rm geom}}^{\mathrm{BTZ}}=\frac{\pi r_{+}}{2G_{N}^{\left(3\right)}}\frac{1}{\mathcal{N}}.
\end{equation}

\noindent The bit-thread flow constructed from the normalized Kalb--Ramond
charge density is therefore

\begin{equation}
v_{\mathrm{BTZ}}^{i}=\frac{1}{4G_{N}^{\left(3\right)}l_{\mathrm{AdS}}}\frac{1}{\cosh\rho}\left(1,0\right).
\end{equation}

\noindent The entropy is obtained by evaluating the flux through the
minimal circle at $\rho=0$. Therefore, the entanglement entropy can
be calculated by the integral:

\begin{equation}
\Phi_{\mathrm{BTZ}}=\int v_{\mathrm{BTZ}}^{i}n_{i}dS=\frac{2\pi r_{+}}{4G_{N}^{\left(3\right)}}=S_{\mathrm{BH}}^{\mathrm{BTZ}}.
\end{equation}

\noindent This is precisely the Bekenstein--Hawking entropy of the
nonrotating BTZ black hole. For the two-sided eternal black hole,
this quantity may equivalently be interpreted as the entanglement
entropy between the two asymptotic boundaries in the thermofield-double
state. In the string theory interpretation, a codimension-2 surface
$\Sigma$, such as an RT surface or a black hole bifurcation surface,
is pierced by the string charge flow lines. The local quantity $v_{\mathrm{BTZ}}^{i}n_{i}dS$
measures the entropy flux/bit-thread flow through an infinitesimal
element of $\Sigma$, while its integral gives the total effective
number of string-charge lines crossing $\Sigma$. 

Moreover, at first sight, the appearance of the horizon radius in
$C_{{\rm geom}}^{\mathrm{BTZ}}$ may seem to make the entropy calculation
tautological. Indeed, because the string charge density is normalized
to carry a total flux $\mathcal{N}$, any proportional maximal entropy
flow must have a conversion coefficient equal to the maximal entropy
flux per unit normalized charge. It is therefore important to state
precisely what is used to determine this coefficient. We do not substitute
the global Bekenstein--Hawking entropy into $C_{\mathrm{geom}}^{\mathrm{BTZ}}$.
Instead, $C_{\mathrm{geom}}^{\mathrm{BTZ}}$ is fixed locally. At
the bifurcation circle, the normalized projected string current has
the density

\begin{equation}
\left|q_{\mathrm{BTZ}}\left(0\right)\right|=\frac{\mathcal{N}}{2\pi r_{+}},
\end{equation}

\noindent whereas the local bit-thread capacity is

\begin{equation}
\left|v_{\mathrm{BTZ}}\left(0\right)\right|=\frac{1}{4G_{N}^{\left(3\right)}}.
\end{equation}

\noindent Requiring the projected string charge density to saturate
this local capacity gives

\begin{equation}
C_{\mathrm{geom}}^{\mathrm{BTZ}}=\frac{\left|v_{\mathrm{BTZ}}\left(0\right)\right|}{\left|q_{\mathrm{BTZ}}\left(0\right)\right|}=\frac{2\pi r_{+}}{4G_{N}^{\left(3\right)}\mathcal{N}}.
\end{equation}
The factor $r_{+}$ appears because the fixed total charge $\mathcal{N}$
is uniformly distributed over a circle of circumference $2\pi r_{+}$.
A larger horizon therefore has a smaller string-charge density for
the same $\mathcal{N}$, and correspondingly requires a larger entropy-per-charge
conversion coefficient. It follows that

\begin{equation}
C_{\mathrm{geom}}^{\mathrm{BTZ}}\mathcal{N}=\frac{2\pi r_{+}}{4G_{N}^{\left(3\right)}}=S_{\mathrm{BH}}^{\mathrm{BTZ}}.
\end{equation}
At this stage, $C_{\mathrm{geom}}^{\mathrm{BTZ}}$ is fixed using
the string-frame norm bound. Therefore, the present calculation shows
that the projected Kalb-Ramond charge density realizes the BTZ maximal
flow and reproduces the Bekenstein-Hawking entropy, but it does not
yet provide a microscopic derivation of the normalization. We return
to this issue in Section 6, where we reverse the logic and ask whether
the corresponding coefficient can instead be fixed from D1-D5 CFT
microscopic counting without imposing the bulk norm bound. This alternative
determination is conditional on the \emph{boundary-matching conjecture}
introduced there. Agreement between the CFT-determined coefficient
and $C_{\mathrm{geom}}$ would then provide a nongeometric consistency
check conditional on the \emph{boundary-matching conjecture}.

The above construction leads to two main motivations for the following
analysis.

\textbf{First}, the discussion above was restricted to three-dimensional
BTZ or locally $\mathrm{AdS}_{3}$ spacetime. In this setting, the
Kalb--Ramond charge density/bit threads correspondence reproduces
both holographic entanglement entropy and the Bekenstein--Hawking
entropy of the BTZ black hole \cite{Wu:2025qwc}. It is therefore
natural to ask whether this correspondence can be generalized to higher-dimensional
backgrounds and whether it can reproduce the Bekenstein--Hawking
entropy of higher-dimensional black holes.

Because the correspondence is formulated in terms of fundamental string
charge, a natural consistent higher-dimensional setting is ten-dimensional
superstring theory rather than an arbitrary higher-dimensional gravitational
theory. Type IIB supergravity admits the F1--NS5--P and D1--D5--P
black string or black brane configurations. After compactification
along their common $S^{1}$ direction and the internal $T^{4}$, these
configurations give five-dimensional black holes. In the decoupling
or near-horizon limit, their geometry contains a locally $\mathrm{BTZ}_{3}$
factor and, in the standard compactification, takes the form \cite{Martinec:2017ztd}

\begin{equation}
\mathrm{BTZ}_{3}\times S^{3}\times T^{4}.
\end{equation}

This structure provides a direct connection with the three-dimensional
construction developed above. It therefore suggests a systematic route
for extending the Kalb--Ramond charge density/bit threads correspondence
to ten-dimensional string backgrounds and for calculating either the
entropy of the ten-dimensional black string configuration or, equivalently,
the entropy of the compactified five-dimensional three-charge black
hole. However, the calculations of the F1--NS5--P and D1--D5--P
black hole entropies using our string-charge method do not require
the near-horizon limit. This shows that our results are not simply
inherited from the BTZ sector and therefore provide a nontrivial higher-dimensional
test of the correspondence.

\begin{figure}[h]
\begin{centering}
\includegraphics[scale=0.35]{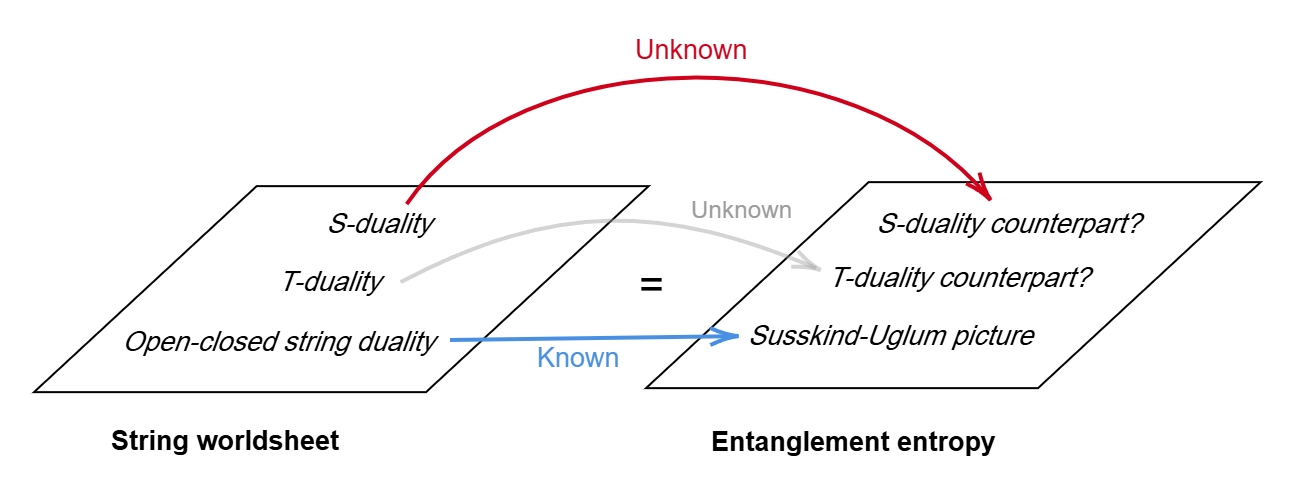}
\par\end{centering}
\caption{\label{fig:duality}This figure illustrates the central idea of this
paper. If the correspondence between string charge density and bit
threads is fundamental, then dualities of string theory should also
have counterparts in terms of entanglement observables. In our previous
work, we proposed that open-closed string duality is related to the
Susskind-Uglum interpretation of black hole entropy. In this paper,
we will focus on the S-duality aspect of this correspondence.}
\end{figure}

\textbf{Second}, the string charge density/bit threads correspondence
suggests that quantities defined in string worldsheet theory may have
counterparts in holographic entanglement theory. As argued in our
previous work, if this correspondence is fundamental, dualities of
string theory should also have corresponding descriptions in terms
of entanglement observables. We previously proposed that open--closed
string duality is related to the Susskind--Uglum interpretation of
black hole entropy. It is therefore natural to investigate whether
additional worldsheet or spacetime dualities can be translated into
relations among entanglement quantities as illustrated in (\ref{fig:duality}).

In particular, the F1--NS5--P and D1--D5--P systems are related
by the S-duality of Type IIB string theory:

\begin{equation}
\mathrm{F1}\longleftrightarrow\mathrm{D1},\qquad\mathrm{NS5}\longleftrightarrow\mathrm{D5},\qquad\mathrm{P}\longleftrightarrow\mathrm{P}.
\end{equation}

\noindent After reproducing the Bekenstein--Hawking entropy in both
duality frames, we can investigate how the string charge density/bit
threads construction transforms under S-duality. Such a comparison
must consistently track the transformations of the metric, dilaton,
two-form gauge fields, string couplings, and compactification data.
Since the physical Bekenstein--Hawking entropy is invariant under
this S-duality, the two kinds of charge density descriptions should
ultimately yield the same entropy, even though their expressions in
the two string frame descriptions may appear different.

\section{Brief review of the F1--NS5--P/D1--D5--P system}

One of the most important results in string theory is the microscopic
interpretation of the Bekenstein--Hawking entropy of certain black
holes. A canonical example is the five-dimensional black hole obtained
from a bound state of fundamental strings (F1), NS5-branes, and momentum
modes (P) along a common compact direction. More precisely, the ten-dimensional
configuration is an F1--NS5--P black string extended along the common
direction $w$. Compactifying the $w$-direction together with the
internal $T^{4}$ gives a five-dimensional black hole, whereas keeping
$w$ noncompact gives a six-dimensional black string after compactification
on $T^{4}$. In the following, we first consider the ten-dimensional
F1--NS5--P configuration.

We begin with the NS--NS sector of Type IIB supergravity in the string
frame. The relevant low-energy effective action is

\begin{equation}
S=\frac{1}{2\kappa_{10}^{2}}\int d^{10}x\sqrt{-g}e^{-2\phi}\left(R+4\left(\nabla\phi\right)^{2}-\frac{1}{12}\left(H^{\left(3\right)}\right)^{2}\right),\label{eq:NSNS action}
\end{equation}

\noindent where $H^{\left(3\right)}=dB^{\left(2\right)}$ , $\left(H^{\left(3\right)}\right)^{2}=H_{\mathrm{abc}}H^{\mathrm{abc}}$
and $2\kappa_{10}^{2}=\left(2\pi\right)^{7}\alpha^{\prime4}$. In
following sections, we set the asymptotic coupling to unity, $g_{s}=1$,
and hence $\phi_{\infty}=0$. Note when $g_{s}$ is restored, the
physical coupling is $2\kappa_{10}^{2}=\left(2\pi\right)^{7}g_{s}^{2}\alpha^{\prime4}$
and $8\pi G_{N}^{\left(10\right)}=\kappa_{10}^{2}$. This convention
avoids double counting the asymptotic dilaton factor. Moreover, all
Ramond--Ramond fields are consistently set to zero in the action
(\ref{eq:NSNS action}). The corresponding equations of motion are

\begin{eqnarray}
R^{\mathrm{ab}}+2\nabla^{\mathrm{a}}\nabla^{\mathrm{b}}\phi-\frac{1}{4}H^{\mathrm{acd}}H_{\:\mathrm{cd}}^{\mathrm{b}} & = & 0,\nonumber \\
\nabla_{\mathrm{a}}\left(e^{-2\phi}H^{\mathrm{abc}}\right) & = & 0,\nonumber \\
4\nabla^{2}\phi-4\left(\nabla\phi\right)^{2}+R-\frac{1}{12}H^{2} & = & 0.
\end{eqnarray}

\noindent The Bianchi identity must also be imposed:

\begin{equation}
dH^{\left(3\right)}=0.
\end{equation}

\noindent To describe the F1--NS5--P system, we decompose the ten-dimensional
spacetime into the following directions:
\begin{itemize}
\item The time direction $t$ and the common string direction $w$.
\item Four noncompact transverse directions $x^{m}$, with $m=1,\ldots,4$,
which form $\mathbb{R}^{4}$. 
\item Four compact directions $y^{i}$, with $i=1,\ldots,4$, which are
taken, for simplicity, to form a flat torus $T^{4}$.
\end{itemize}
The rotational symmetry of the transverse $\mathbb{R}^{4}$ is $SO\left(4\right)$.
Introducing the radial coordinate

\begin{equation}
r\equiv\sqrt{\left(x^{1}\right)^{2}+\cdots+\left(x^{4}\right)^{2}},
\end{equation}

\noindent the flat transverse metric can be written as

\begin{equation}
\delta_{mn}dx^{m}dx^{n}=dr^{2}+r^{2}d\Omega_{3}^{2}.
\end{equation}

\noindent Rotational invariance implies that all nontrivial functions
depend only on $r$. For simplicity, we avoid introducing nonextremal
boost parameters and restrict attention to the extremal BPS F1--NS5--P
black string and use the ansatz

\begin{eqnarray}
dS_{\mathrm{F1-NS5-P}}^{2} & = & \frac{1}{H_{1}}\left[-dt^{2}+dw^{2}+K\left(r\right)\left(dt+dw\right)^{2}\right]+H_{5}\left(dr^{2}+r^{2}d\Omega_{3}^{2}\right)+\delta_{ij}dy^{i}dy^{j},\nonumber \\
\phi & = & \phi\left(r\right),\qquad H=\frac{1}{2}\partial_{r}\alpha\left(r\right)dr\wedge dt\wedge dw+2Q_{5}\omega_{3}.
\end{eqnarray}

\noindent The term $K\left(r\right)\left(dt+dw\right)^{2}$ describes
a chiral pp-wave propagating along the common F1--NS5 direction.
Its null character is appropriate for the BPS momentum charge. Here
we restrict to the extremal F1--NS5--P solution. A general nonextremal
solution contains additional functions and parameters that are not
included in our ansatz.

The functions $H_{1}\left(r\right)$, $H_{5}\left(r\right)$, $K\left(r\right)$
and $\alpha\left(r\right)$ are to be determined from the equations
of motion. The Bianchi identity $dH=0$ is automatically satisfied
away from the location of the branes because $d\omega_{3}=0$. The
two-form potential for the magnetic part can only be defined patchwise
on $S^{3}$, since the nonzero magnetic flux prevents the existence
of a globally defined $\omega_{2}$.

Substituting this ansatz into the equations of motion and choosing
$\alpha\left(r\right)=2\left(1-H_{1}^{-1}\right)$ to reproduce the
electric F1 field gives $\frac{1}{2}\partial_{r}\alpha\left(r\right)=\partial_{r}\left(1-H_{1}^{-1}\right)$.
The dilaton is fixed by $e^{2\phi}=H_{5}/H_{1}$, with the asymptotic
string coupling set to unity. The remaining equations reduce to the
harmonic equations

\begin{equation}
\nabla_{\mathbb{R}^{4}}^{2}H_{1}=0,\qquad\nabla_{\mathbb{R}^{4}}^{2}H_{5}=0,\qquad\nabla_{\mathbb{R}^{4}}^{2}K=0,
\end{equation}

\noindent away from $r=0$. For a spherically symmetric function $F\left(r\right)$,
the flat Laplacian on the transverse $\mathbb{R}^{4}$ is 

\begin{equation}
\nabla_{\mathbb{R}^{4}}^{2}F\left(r\right)=\frac{1}{r^{3}}\partial_{r}\left(r^{3}\partial_{r}F\right).
\end{equation}

\noindent Therefore, 

\begin{equation}
\nabla_{\mathbb{R}^{4}}^{2}F\left(r\right)=0,
\end{equation}
implies 

\begin{equation}
F\left(r\right)=c_{0}+\frac{c_{1}}{r^{2}}.
\end{equation}

\noindent The solutions that approach ten-dimensional Minkowski spacetime
as $r\rightarrow\infty$ are consequently \cite{Tseytlin:1996bh,Mathur:2005zp,Martinec:2017ztd,Martinec:2018nco}

\begin{equation}
H_{1}\left(r\right)=1+\frac{Q_{1}}{r^{2}},\qquad H_{5}\left(r\right)=1+\frac{Q_{5}}{r^{2}},\qquad K\left(r\right)=\frac{Q_{P}}{r^{2}}.
\end{equation}

\noindent The constants $Q_{1}$, $Q_{5}$ and $Q_{P}$ are proportional,
after the compactification parameters and normalization factors are
included, to the numbers of F1 strings, NS5-branes, and momentum quanta,
respectively.

The Kalb--Ramond potential can be chosen locally as

\begin{equation}
B^{\left(2\right)}=\left(1-H_{1}^{-1}\right)dt\wedge dw+2Q_{5}\omega_{2},
\end{equation}

\noindent where $d\omega_{2}=\omega_{3}$. Its field strength is

\begin{equation}
H^{\left(3\right)}=d\left(1-H_{1}^{-1}\right)\wedge dt\wedge dw+2Q_{5}\omega_{3}
\end{equation}

\noindent Here $\omega_{3}$ is the volume form of the unit three-sphere.
In angular coordinates, it may be written as

\begin{equation}
\omega_{3}=\sin^{2}\chi\sin\vartheta d\chi\wedge d\vartheta\wedge d\varphi,
\end{equation}

\noindent with the normalization $\int_{S^{3}}\omega_{3}=2\pi^{2}$.
The electric component of $H$ carries the F1 charge, while the magnetic
component carries the NS5 charge. Notice that the F1 strings, NS5-branes,
and momentum wave are all coincident at the origin $r=0$ of the transverse
$\mathbb{R}^{4}$. The F1 strings extend along $w$, the NS5-branes
wrap $w$ and the entire $T^{4}$, and the pp-wave component $K\left(r\right)$
supplies the momentum along $w$, see Table (\ref{tab:table}).

\begin{table}[h]
\centering{}%
\begin{tabular}{c|cccc}
 & $t$ & $w$ & $T^{4}$ & $\mathbb{R}_{\bot}^{4}$\tabularnewline
\hline 
F1 & $\times$ & $\times$ & $-$ & $-$\tabularnewline
NS5 & $\times$ & $\times$ & $\times$ & $-$\tabularnewline
P & $\times$ & $\rightarrow$ & $-$ & $-$\tabularnewline
\end{tabular}\caption{\label{tab:table}The solution describes fundamental strings extended
along $w$, NS5-branes wrapped on $w\times T^{4}$, and a right-moving
momentum wave along $w$. The electric component of $H$ carries $Q_{1}$,
the magnetic flux through $S^{3}$ carries $Q_{5}$, and the null
metric deformation $K$ carries $Q_{P}$.}
\end{table}

The solution satisfies the source-free equations of motion for $r>0$.
The charges are read off from flux integrals rather than from explicit
delta function sources. For example,

\begin{equation}
Q_{1}\propto\int_{S^{3}\times T^{4}}e^{-2\phi}\star H,
\end{equation}

\noindent whereas

\begin{equation}
Q_{5}\propto\int_{S^{3}}H.
\end{equation}

\noindent %

\subsection{Near-horizon limit}

We now consider the near-horizon limit of the extremal F1--NS5--P
black-string solution. For the extremal geometry, the horizon is located
at

\begin{equation}
r=0.
\end{equation}

\noindent The throat region is obtained by taking

\begin{equation}
r^{2}\ll Q_{1},Q_{5}.
\end{equation}

\noindent In this region, the harmonic functions become

\begin{equation}
H_{1}\simeq\frac{Q_{1}}{r^{2}},\qquad H_{5}\simeq\frac{Q_{5}}{r^{2}},\qquad K\simeq\frac{Q_{P}}{r^{2}}.
\end{equation}

\noindent Substituting these expressions into the ten-dimensional
solution gives \cite{Martinec:2017ztd}

\begin{equation}
dS_{\mathrm{F1-NS5-P}}^{2}=\frac{r^{2}}{Q_{1}}\left[-dt^{2}+dw^{2}+\frac{Q_{P}}{r^{2}}\left(dt+dw\right)^{2}\right]+\frac{Q_{5}}{r^{2}}\left(dr^{2}+r^{2}d\Omega_{3}^{2}\right)+\delta_{ij}dy^{i}dy^{j},
\end{equation}

\noindent The near-horizon dilaton is constant:

\begin{equation}
e^{2\phi}=\frac{Q_{5}}{Q_{1}}=\mathrm{constant}.
\end{equation}

\noindent Using the following coordinate transformation

\begin{equation}
z=\frac{\sqrt{Q_{1}Q_{5}}}{r},\qquad\frac{dr^{2}}{r^{2}}=\frac{dz^{2}}{z^{2}}.
\end{equation}

\noindent The three-dimensional part of the near-horizon metric then
takes the form

\begin{equation}
dS_{3}^{2}=\frac{Q_{5}}{z^{2}}\left(-dt^{2}+dw^{2}+dz^{2}\right)+\frac{Q_{P}}{Q_{1}}\left(dt+dw\right)^{2}.
\end{equation}

\noindent This is the extremal BTZ geometry written in a null-wave
coordinate system, provided that the spatial direction is periodically
identified. The momentum term selects a null direction on the AdS$_{3}$
boundary and generates the extremal BTZ quotient. Because three-dimensional
Einstein gravity has no local propagating gravitational degrees of
freedom, the resulting geometry remains locally AdS$_{3}$. The corresponding
AdS radius can be identified as

\begin{equation}
l_{\mathrm{AdS}}^{2}=Q_{5}.
\end{equation}

\noindent The extremal horizon is located at $r=0$, which corresponds
to $z\rightarrow\infty$. It is therefore a degenerate horizon rather
than a finite $z$ nonextremal horizon. If the momentum charge is
set to zero, $Q_{P}=0$, the three-dimensional local metric reduces
to the Poincare patch of AdS$_{3}$:

\begin{equation}
dS_{3}^{2}=\frac{Q_{5}}{z^{2}}\left(-dt^{2}+dw^{2}+dz^{2}\right).
\end{equation}

\noindent Thus, for nonzero momentum and a compact spatial identification,
the three-dimensional geometry is extremal BTZ. On the other hand,
when the momentum vanishes, it reduces to the Poincare patch of pure
AdS$_{3}$. Therefore, the full near-horizon geometry can be interpreted
as $\mathrm{BTZ}_{3}\times S^{3}\times T^{4}$ for $Q_{P}\neq0$,
and as $\mathrm{AdS}_{3}\times S^{3}\times T^{4}$ for $Q_{P}=0$.

To understand what happens physically in the near-horizon limit, we
can trace the scale $Q_{5}$, which determines both the radius of
$S^{3}$ and the cosmological constant of the AdS$_{3}$ factor in
the F1--NS5-P solution. Specifically, after dropping the additive
constants in $H_{1}$ and $H_{5}$, the asymptotically flat region
is removed and the geometry reduces to the decoupled near-horizon
region. The F1 and NS5 charges remain encoded in the background fluxes,
which determine the size of the internal space, the constant dilaton,
and the curvature scale of the AdS$_{3}$ factor. The corresponding
microscopic degrees of freedom are described by a two-dimensional
CFT living on the boundary of AdS$_{3}$.

\subsection{From F1--NS5--P to D1--D5--P via S-duality}

We now apply the S-duality of Type IIB string theory to obtain the
D1--D5--P solution. Type IIB supergravity possesses an $SL\left(2,\mathbb{Z}\right)$
symmetry that acts on the complex field

\begin{equation}
\tau=C^{\left(0\right)}+ie^{-\phi},
\end{equation}

\noindent where $C^{\left(0\right)}$ is the R--R $0$\nobreakdash-form.
A general $SL\left(2,\mathbb{Z}\right)$ transformation is given by

\begin{equation}
\tau\rightarrow\tau^{\prime}=\frac{a\tau+b}{c\tau+d},\qquad\left(\begin{array}{cc}
a & b\\
c & d
\end{array}\right)\in SL\left(2,\mathbb{Z}\right).
\end{equation}

\noindent The matrix representing the S-transformation can be chosen
as

\begin{equation}
S=\left(\begin{array}{cc}
0 & -1\\
1 & 0
\end{array}\right),
\end{equation}
which gives

\begin{equation}
\tau\rightarrow\tau^{\prime}=-\frac{1}{\tau}.
\end{equation}

\noindent For the F1--NS5--P solution,

\begin{equation}
C^{\left(0\right)}=0,\qquad\tau=ie^{-\phi}.
\end{equation}

\noindent Using S-duality, we obtain

\begin{equation}
\tau^{\prime}=-\frac{1}{\tau}=-\frac{1}{ie^{-\phi}}=ie^{\phi}.
\end{equation}

\noindent Comparing this with the definition in the D1--D5--P duality
frame,

\begin{equation}
\tau^{\prime}=C^{\left(0\right)\prime}+ie^{-\phi^{\prime}},
\end{equation}

\noindent we find

\begin{equation}
C^{\left(0\right)\prime}=0,\qquad e^{-\phi^{\prime}}=e^{\phi}.
\end{equation}

\noindent Therefore,

\begin{equation}
\phi^{\prime}=-\phi,\qquad e^{2\phi^{\prime}}=e^{-2\phi}=\frac{H_{1}}{H_{5}}.
\end{equation}

\noindent Here and below, $\phi^{\prime}$ denotes the D1-D5-P frame
dilaton shifted by its asymptotic value, so that $\phi_{\infty}^{\prime}=0$.
On the other hand, the S-duality transformation also exchanges the
NS--NS and R--R two-form potentials. With the convention

\begin{equation}
\left(\begin{array}{c}
B^{\left(2\right)\prime}\\
C^{\left(2\right)\prime}
\end{array}\right)=\left(\begin{array}{cc}
0 & -1\\
1 & 0
\end{array}\right)\left(\begin{array}{c}
B^{\left(2\right)}\\
C^{\left(2\right)}
\end{array}\right).
\end{equation}

\noindent We obtain

\begin{equation}
B^{\left(2\right)\prime}=-C^{\left(2\right)},\qquad C^{\left(2\right)\prime}=B^{\left(2\right)}.
\end{equation}

\noindent Since the initial F1--NS5--P solution contains only the
Kalb--Ramond field and has

\begin{equation}
C^{\left(0\right)}=C^{\left(2\right)}=C^{\left(4\right)}=0,
\end{equation}

\noindent the transformed background satisfies

\begin{equation}
B^{\left(2\right)\prime}=0,\qquad C^{\left(2\right)\prime}=B^{\left(2\right)}.
\end{equation}

\noindent The R--R three-form field strength is consequently

\begin{equation}
F^{\left(3\right)}=dC^{\left(2\right)\prime}=d\left(1-H_{1}^{-1}\right)\wedge dt\wedge dw+2Q_{5}\omega_{3}.
\end{equation}

\noindent Under this S-duality transformation, the fundamental string,
which couples electrically to $B$, is mapped to an object that couples
electrically to $C^{\left(2\right)\prime}$, namely the D1-brane.
The NS5-brane is the magnetic source for $B$. After S-duality, this
magnetic NS--NS flux becomes R--R flux, whose magnetic source is
the D5-brane. The momentum charge is carried by the metric and is
neutral under S-duality. Consequently, the pp-wave term remains unchanged.
Therefore,

\begin{equation}
\mathrm{\left(F1,NS5,P\right)}\qquad\xrightarrow{\quad S\quad}\qquad\mathrm{\left(D1,D5,P\right)}.
\end{equation}

To avoid confusion between different conformal frames, we work temporarily
in the Einstein frame, where the S-duality transformation is simplest.
The Einstein-frame metric $g_{\mathrm{E}}$ is related to the string-frame
metric $g_{\mathrm{string}}$ by

\begin{equation}
g_{\mathrm{E}}=e^{-\phi/2}g_{\mathrm{string}},\qquad dS_{\mathrm{E}}^{2}=e^{-\phi/2}dS_{\mathrm{string}}^{2}.
\end{equation}

\noindent For the F1--NS5--P solution, $e^{-\phi/2}=\left(H_{1}/H_{5}\right)^{1/4}$.
Therefore, the Einstein-frame metric is

\begin{equation}
dS_{\mathrm{F1-NS5-P}}^{2}=\left(\frac{H_{1}}{H_{5}}\right)^{1/4}\left(\frac{1}{H_{1}}\left[-dt^{2}+dw^{2}+K\left(r\right)\left(dt+dw\right)^{2}\right]+H_{5}\left(dr^{2}+r^{2}d\Omega_{3}^{2}\right)+\delta_{ij}dy^{i}dy^{j}\right).
\end{equation}

\noindent The Einstein-frame metric is invariant under the Type IIB
S-duality transformation, since the initial solution has $C^{\left(0\right)}=0$,
the transformation acts as

\begin{equation}
\phi^{\prime}=-\phi,\qquad g_{E}^{\prime}=g_{E},\qquad B^{\left(2\right)\prime}=-C^{\left(2\right)}=0,\qquad C^{\left(2\right)\prime}=B^{\left(2\right)}.
\end{equation}

\noindent The new dilaton satisfies

\begin{equation}
e^{\phi^{\prime}/2}=e^{-\phi/2}=\left(\frac{H_{1}}{H_{5}}\right)^{1/4}.
\end{equation}

\noindent Now, we continue the calculation using the D1--D5--P string-frame
metric, which is obtained from the

\begin{equation}
g_{\mathrm{string}}^{\prime}=e^{\phi^{\prime}/2}g_{\mathrm{E}}^{\prime}.
\end{equation}

\noindent Equivalently

\begin{equation}
g_{\mathrm{string}}^{\prime}=e^{-\phi}g_{\mathrm{string}}=\sqrt{\frac{H_{1}}{H_{5}}}g_{\mathrm{string}}.
\end{equation}

\noindent We therefore have

\begin{eqnarray}
dS_{\mathrm{D1-D5-P}}^{2} & = & \frac{1}{\sqrt{H_{1}H_{5}}}\left[-dt^{2}+dw^{2}+K\left(r\right)\left(dt+dw\right)^{2}\right]\nonumber \\
 &  & +\sqrt{H_{1}H_{5}}\left(dr^{2}+r^{2}d\Omega_{3}^{2}\right)+\sqrt{\frac{H_{1}}{H_{5}}}\delta_{ij}dy^{i}dy^{j},\nonumber \\
e^{2\phi^{\prime}} & = & \frac{H_{1}}{H_{5}},\qquad C^{\left(2\right)\prime}=\left(1-H_{1}^{-1}\right)dt\wedge dw+2Q_{5}\omega_{2}.
\end{eqnarray}

\noindent This is the extremal D1--D5--P solution \cite{Cvetic:1995uj,Tseytlin:1996as,Horowitz:1996ay,Giusto:2010gv}.
In particular,

\begin{equation}
H_{1}\left(r\right)=1+\frac{Q_{1}}{r^{2}},\qquad H_{5}\left(r\right)=1+\frac{Q_{5}}{r^{2}},\qquad K\left(r\right)=\frac{Q_{P}}{r^{2}}.
\end{equation}

\noindent In the near-horizon limit, the D1--D5--P geometry is again
$\mathrm{BTZ}_{3}\times S^{3}\times T^{4}$. In the D1--D5--P string
frame, the AdS radius is $l_{\mathrm{AdS}}^{2}=\sqrt{Q_{1}Q_{5}}$,
whereas in the F1--NS5-P string frame, $l_{\mathrm{AdS}}^{2}=Q_{5}$.
This difference is simply due to the S-duality rescaling of the string
frame metric. The two string frame metrics therefore assign different
length scales, while the Einstein-frame geometry is unchanged. Consequently,
the physical black hole entropy is the same in both frames.

On the other hand, the corresponding action after S-duality is \cite{Maldacena:1998bw}

\begin{equation}
S^{\prime}=\frac{1}{2\kappa_{10}^{2}}\int d^{10}x\sqrt{-g^{\prime}}\left[e^{-2\phi^{\prime}}\left(R^{\prime}+4\left(\nabla^{\prime}\phi^{\prime}\right)^{2}\right)-\frac{1}{12}\left(F^{\left(3\right)}\right)^{2}\right],
\end{equation}

\noindent where $F_{\mathrm{abc}}=dC^{\left(2\right)\prime}=3\partial_{\left[\mathrm{a}\right.}C_{\left.\mathrm{bc}\right]}^{\left(2\right)\prime}$.
Since the original action contains only the Kalb--Ramond field, the
transformed background satisfies

\begin{eqnarray*}
 &  & B^{\left(2\right)\prime}=0,\qquad H^{\left(3\right)\prime}=0,\\
 &  & C^{\left(0\right)\prime}=0,\qquad F^{\left(1\right)}=dC^{\left(0\right)\prime}=0,
\end{eqnarray*}

\noindent and

\begin{equation}
F^{\left(3\right)}=dC^{\left(2\right)\prime}-C^{\left(0\right)\prime}H^{\left(3\right)\prime}=dC^{\left(2\right)\prime}.
\end{equation}

\noindent Moreover $C^{\left(4\right)\prime}=0$, and hence

\begin{equation}
F^{\left(5\right)}=dC^{\left(4\right)\prime}-\frac{1}{2}C^{\left(2\right)\prime}\wedge H^{\left(3\right)\prime}+\frac{1}{2}B^{\left(2\right)\prime}\wedge dC^{\left(2\right)\prime}=0.
\end{equation}

\noindent The equations of motion obtained from the string-frame action
are

\begin{eqnarray}
R^{\prime\mathrm{ab}}+2\nabla^{\prime\mathrm{a}}\nabla^{\prime\mathrm{b}}\phi^{\prime}-\frac{1}{4}e^{2\phi^{\prime}}\left(F^{\mathrm{acd}}F_{\:\mathrm{cd}}^{\mathrm{b}}-\frac{1}{6}g^{\prime\mathrm{ab}}F^{\mathrm{cde}}F_{\mathrm{cde}}\right) & = & 0,\nonumber \\
\nabla_{\mathrm{a}}^{\prime}F^{\mathrm{abc}} & = & 0,\nonumber \\
4\nabla^{\prime2}\phi^{\prime}-4\left(\nabla^{\prime}\phi^{\prime}\right)^{2}+R^{\prime} & = & 0.
\end{eqnarray}

\noindent These equations must be supplemented by the R--R Bianchi
identity

\begin{equation}
dF^{\left(3\right)}=0.
\end{equation}

\noindent In the metric equation, the NS--NS contribution $-\frac{1}{4}H^{\mathrm{acd}}H_{\:\mathrm{cd}}^{\mathrm{b}}$
of F1-NS5-P configuration is replaced by the R--R contribution $-\frac{1}{4}e^{2\phi^{\prime}}\left(F^{\mathrm{acd}}F_{\:\mathrm{cd}}^{\mathrm{b}}-\frac{1}{6}g^{\mathrm{ab}}F^{\mathrm{cde}}F_{\mathrm{cde}}\right)$.
Because the R--R kinetic term is not multiplied by $e^{-2\phi}$
in the string-frame action, its variation with respect to the metric
produces the additional trace term proportional to $\frac{1}{24}e^{2\phi^{\prime}}g^{\mathrm{ab}}F^{2}$
inside the complete metric equation. By contrast, the R--R three-form
does not appear explicitly in the dilaton equation obtained by directly
varying the dilaton. The R--R field nevertheless affects the dilaton
indirectly through the coupled metric equations. The F1--NS5--P
and D1--D5--P backgrounds thus provide two S-dual descriptions of
the same extremal system. Their string-frame metrics, dilatons, and
two-form fields are different, but their Einstein-frame geometry and
physical Bekenstein--Hawking entropy agree. This makes the pair particularly
suitable for testing how the string charge density/bit threads correspondence
transforms under S-duality.

\section{F1-NS5-P with a macroscopic F1 string source}

Because our previous work was based on the correspondence between
the Kalb--Ramond charge density and bit threads, a natural ten-dimensional
generalization is to consider the NS--NS sector of Type IIB supergravity.
In the string frame, the relevant truncated action is

\begin{equation}
S=\frac{1}{2\kappa_{10}^{2}}\int d^{10}x\sqrt{-g}e^{-2\phi}\left(R+4\left(\nabla\phi\right)^{2}-\frac{1}{12}\left(H^{\left(3\right)}\right)^{2}\right)+S_{\mathrm{F1}},
\end{equation}

\noindent where $2\kappa_{10}^{2}=\left(2\pi\right)^{7}\alpha^{\prime4}$.
Here we set the asymptotic string coupling to unity, $g_{s}=1$, so
that $\phi_{\infty}=0$, as before. A fundamental string, known as
F1, couples electrically to the Kalb--Ramond two-form $B$. A macroscopic
F1-string source is described by the Polyakov action

\begin{equation}
S_{\mathrm{F1}}=-\frac{1}{4\pi\alpha^{\prime}}\int d^{2}\sigma\left(\sqrt{-\gamma}\gamma^{mn}\partial_{m}X^{\mathrm{a}}\partial_{n}X^{\mathrm{b}}g_{\mathrm{ab}}+\epsilon^{mn}\partial_{m}X^{\mathrm{a}}\partial_{n}X^{\mathrm{b}}B_{\mathrm{ab}}\right),
\end{equation}

\noindent The bulk equations of motion are 

\begin{eqnarray}
R^{\mathrm{ab}}+2\nabla^{\mathrm{a}}\nabla^{\mathrm{b}}\phi-\frac{1}{4}H^{\mathrm{acd}}H_{\:\mathrm{cd}}^{\mathrm{b}} & = & \kappa_{10}^{2}T_{\mathrm{F1}}^{\mathrm{ab}},\nonumber \\
\nabla_{\mathrm{a}}\left(e^{-2\phi}H^{\mathrm{abc}}\right) & = & \frac{\kappa_{10}^{2}}{\pi\alpha^{\prime}}j_{\mathrm{F1}}^{\mathrm{bc}}\nonumber \\
4\nabla^{2}\phi-4\left(\nabla\phi\right)^{2}+R-\frac{1}{12}H^{2} & = & 0,
\end{eqnarray}

\noindent where the stress tensor and the antisymmetric string current
are

\begin{eqnarray}
T_{\mathrm{F1}}^{\mathrm{ab}} & = & -\frac{e^{2\phi}}{2\pi\alpha^{\prime}\sqrt{-g}}\int d^{2}\sigma\sqrt{-\gamma}\gamma^{mn}\partial_{m}X^{\mathrm{a}}\partial_{n}X^{\mathrm{b}}\delta^{\left(10\right)}\left(x-X\left(\tau,\sigma\right)\right),\nonumber \\
j_{\mathrm{F1}}^{\mathrm{ab}} & = & \frac{1}{\sqrt{-g}}\int d^{2}\sigma\left(\frac{\partial X^{\mathrm{a}}}{\partial\tau}\frac{\partial X^{\mathrm{b}}}{\partial\sigma}-\frac{\partial X^{\mathrm{b}}}{\partial\tau}\frac{\partial X^{\mathrm{a}}}{\partial\sigma}\right)\delta^{\left(10\right)}\left(x-X\left(\tau,\sigma\right)\right).\label{eq: source-2-1}
\end{eqnarray}

\noindent Unlike the three-dimensional noncritical string action considered
previously, this ten-dimensional critical Type IIB action contains
no cosmological constant. We treat the additional radial strings as
probes. For a single source, if $L$ denotes the curvature scale,
one may simultaneously require 
\begin{equation}
\epsilon_{{\rm probe}}\equiv\frac{G_{N}^{\left(10\right)}T_{{\rm probe}}}{L^{6}}\ll1.\label{eq:probelimit}
\end{equation}

\noindent Under this condition, the source backreaction is subleading,
and the zeroth-order spacetime remains the source-free F1--NS5--P
black string background. If we consider finite source coupling, the
radial source must be included in the coupled field equations, and
the unperturbed F1--NS5--P geometry is no longer an exact solution.
Therefore, the extremal F1-NS5-P solution in the string frame is

\begin{eqnarray}
dS_{\mathrm{F1-NS5-P}}^{2} & = & \frac{1}{H_{1}}\left[-dt^{2}+dw^{2}+K\left(dt+dw\right)^{2}\right]+H_{5}\left(dr^{2}+r^{2}d\Omega_{3}^{2}\right)+\delta_{ij}dy^{i}dy^{j},\nonumber \\
e^{2\phi\left(r\right)} & = & \frac{H_{5}}{H_{1}},\qquad H=d\left(1-H_{1}^{-1}\right)\wedge dt\wedge dw+2Q_{5}\omega_{3},
\end{eqnarray}

\noindent with

\begin{equation}
H_{1}\left(r\right)=1+\frac{Q_{1}}{r^{2}},\qquad H_{5}\left(r\right)=1+\frac{Q_{5}}{r^{2}},\qquad K\left(r\right)=\frac{Q_{P}}{r^{2}}.
\end{equation}

\noindent Here $\omega_{3}$ is the volume form of the unit three-sphere.
Before specifying the radial probe, one must take account of the momentum-induced
shift in the $w$ direction. The $\left(t,w\right)$ part of the F1--NS5--P
metric can be written as

\begin{equation}
dS_{tw}^{2}=H_{1}^{-1}\left[\left(K-1\right)dt^{2}+2Kdtdw+\left(1+K\right)dw^{2}\right].\label{eq:F1twsector}
\end{equation}

\noindent Comparing this expression with the ADM form

\begin{equation}
dS^{2}=-N_{\mathrm{ADM}}^{2}dt^{2}+h_{ww}\left(dw+\beta^{w}dt\right)^{2}+h_{rr}dr^{2}+\cdots,\label{eq:F1ADMform}
\end{equation}

\noindent gives

\begin{equation}
h_{ww}=\frac{1+K}{H_{1}},\qquad\beta^{w}=\frac{K}{1+K},\qquad N_{\mathrm{ADM}}^{2}=\frac{1}{H_{1}\left(1+K\right)},\qquad h_{rr}=H_{5}.\label{eq:F1ADMdata}
\end{equation}

\noindent A strictly static embedding with $X^{w}=w_{0}$ has the
tangent $\partial_{\tau}X=\partial_{t}$ and hence the induced component

\begin{equation}
\gamma_{\tau\tau}^{\mathrm{ind}}=g_{tt}=\frac{K-1}{H_{1}}.\label{eq:F1staticinduced}
\end{equation}

\noindent Because $K=Q_{P}/r^{2}$ grows toward the extremal horizon,
this component becomes positive when $K>1$. The fixed-$w$ worldsheet
would then cease to be timelike. This reflects the fact that the coordinate
vector $\partial_{t}$ is not aligned with the ADM normal in the presence
of the momentum-induced shift. Thus the fixed-$w$ embedding is not
a suitable probe worldsheet near the horizon.

Since the bit-thread construction is defined on the constant-time
slice $t=0$, it is sufficient to choose an embedding adapted to the
ADM normal on that slice:

\begin{equation}
X^{t}\left(\tau,\sigma\right)=\tau,\qquad X^{r}\left(\tau,\sigma\right)=\sigma,\qquad X^{w}\left(\tau,\sigma\right)=w_{0}-\beta^{w}\left(\sigma\right)\tau,\label{eq:F1sliceembedding}
\end{equation}

\noindent with $X^{y^{i}}=y_{0}^{i}$ and $X^{\Omega^{\alpha}}=\Omega_{0}^{\alpha}$.
At $\tau=0$, its two tangent vectors are

\begin{equation}
e_{\tau}^{\mu}=\partial_{\tau}X^{\mu}=\left(\partial_{t}-\beta^{w}\partial_{w}\right)^{\mu}=N_{\mathrm{ADM}}n^{\mu},\qquad e_{\sigma}^{\mu}=\partial_{\sigma}X^{\mu}=\left(\partial_{r}\right)^{\mu},\label{eq:F1tangents}
\end{equation}

\noindent where $n^{\mu}=N_{\mathrm{ADM}}^{-1}\left(\partial_{t}-\beta^{w}\partial_{w}\right)^{\mu}$
is the future-directed unit normal to the constant-time slice. The
induced metric at $\tau=0$ therefore has components

\begin{equation}
\gamma_{\tau\tau}^{\mathrm{ind}}=-N_{\mathrm{ADM}}^{2}=-\frac{1}{H_{1}\left(1+K\right)},\qquad\gamma_{\tau\sigma}^{\mathrm{ind}}=0,\qquad\gamma_{\sigma\sigma}^{\mathrm{ind}}=H_{5},\label{eq:F1adaptedinduced}
\end{equation}

\noindent so that $\det\gamma^{\mathrm{ind}}=-H_{5}/\left[H_{1}\left(1+K\right)\right]<0$.
The adapted worldsheet is therefore Lorentzian at every exterior point
$r>0$ as the horizon is approached, and its spatial tangent lies
purely in the radial direction on the selected slice. Away from $\tau=0$
one has $e_{\sigma}^{w}=-\tau\partial_{r}\beta^{w}$; this does not
affect the current pulled back to the slice used in the entropy calculation.
This prescription fixes only the slice data needed below; it is not
a claim that the displayed embedding is a stationary worldsheet solution
away from $\tau=0$.

The antisymmetric numerator of the string current is $\mathcal{J}^{\mu\nu}=e_{\tau}^{\mu}e_{\sigma}^{\nu}-e_{\tau}^{\nu}e_{\sigma}^{\mu}$.
At $\tau=0$,

\begin{equation}
\mathcal{J}^{0r}=1,\qquad\mathcal{J}^{0w}=0,\qquad\mathcal{J}^{wr}=-\beta^{w}.\label{eq:F1currentcomponents}
\end{equation}

\noindent At $t=0$, $j_{\mathrm{F1}}^{0w}=0$, whereas $j_{\mathrm{F1}}^{0r}$
is identical to the radial component obtained from the formal fixed-$w$
embedding. Away from this slice, $j_{\mathrm{F1}}^{wr}$ is generally
nonzero. The moving support together with the $0w$ and $wr$ components
ensures the covariant conservation of the full spacetime current,
$\partial_{\mu}\left(\sqrt{-g}j^{\mu\nu}\right)=0$. Consequently,
$q^{w}=0$ and the projected string charge density, and hence the
bit-thread flow used below, is purely radial on $t=0$ without relying
on a potentially spacelike static probe.

The unit-three-sphere metric is

\begin{equation}
d\Omega_{3}^{2}=\bar{\gamma}_{\alpha\beta}d\Omega^{\alpha}d\Omega^{\beta}=d\chi^{2}+\sin^{2}\chi d\theta^{2}+\sin^{2}\chi\sin^{2}\theta d\varphi^{2}.
\end{equation}

\noindent We use $\bar{\gamma}_{\alpha\beta}$ for the unit-sphere
metric in order to distinguish it from the worldsheet metric $\gamma_{mn}$.
The covariant delta function on $S^{3}$ is normalized by

\begin{equation}
\int_{S^{3}}d^{3}\Omega\sqrt{\bar{\gamma}\left(\Omega\right)}\delta_{S^{3}}\left(\Omega,\Omega_{0}\right)=1.
\end{equation}

\noindent After performing the $\tau$ and $\sigma$ integrations,
the nonvanishing component of the antisymmetric current (\ref{eq: source-2-1})
is

\begin{equation}
j_{\mathrm{F1}}^{0r}=\frac{\sqrt{\bar{\gamma}}}{\sqrt{-g}}\delta\left(w-w_{0}\right)\delta_{T^{4}}^{\left(4\right)}\left(y-y_{0}\right)\delta_{S^{3}}\left(\Omega,\Omega_{0}\right),
\end{equation}

\noindent where the factor $\sqrt{\bar{\gamma}}$ comes from rewriting
the coordinate delta function on $S^{3}$ as a covariant delta function.
Note that this radial worldsheet is an auxiliary, anchored probe,
not one of the F1 strings that generates the background charge $Q_{1}$. 

\begin{figure}[h]
\begin{centering}
\includegraphics[scale=0.25]{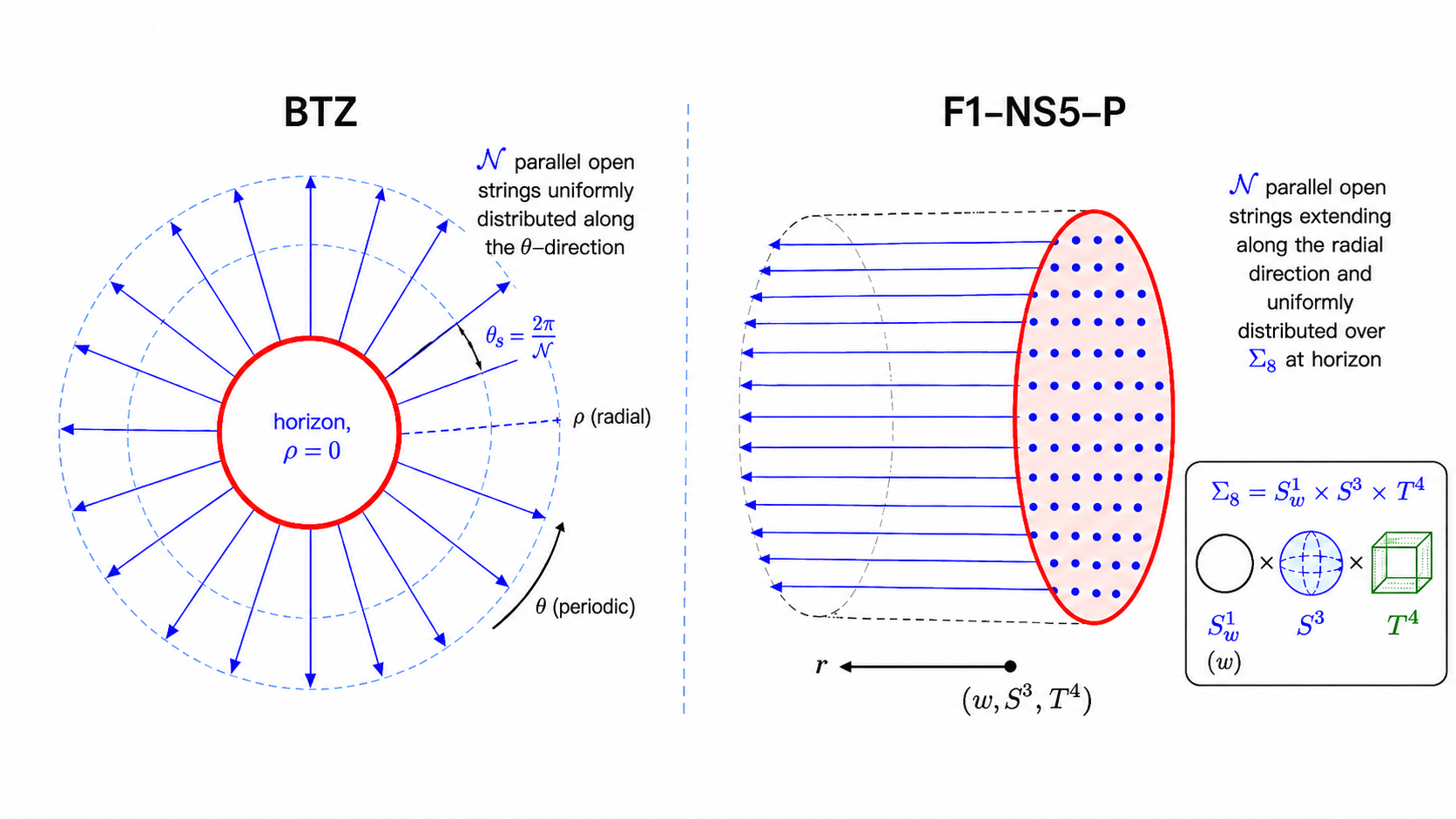}
\par\end{centering}
\caption{\label{fig:Nstring}The left panel illustrates that $\mathcal{N}$
parallel macroscopic strings are uniformly distributed along the angular
direction $\theta$. Each string pierces the horizon orthogonally
and occupies a one-dimensional angular cell of size $\theta_{s}=2\pi/\mathcal{N}$.
The right panel illustrates that $\mathcal{N}$ parallel macroscopic
strings are uniformly distributed over the compact transverse section
$\Sigma_{8}=S_{w}^{1}\times S^{3}\times T^{4}$ at the horizon. Each
string pierces the horizon orthogonally and occupies an eight-dimensional
cell of volume $\Delta\mathcal{V}_{0}=\mathcal{V}_{0}/\mathcal{N}$,
as shown by the blue dots in the figure.}
\end{figure}

We now explain how the three-dimensional construction can be generalized.
In the BTZ example, we considered $\mathcal{N}$ parallel radial strings
distributed uniformly along the horizon circle. Their angular separation
was

\begin{equation}
\theta_{s}=\frac{2\pi}{\mathcal{N}},
\end{equation}

\noindent so that each string pierced the horizon orthogonally. In
a higher-dimensional black hole, the horizon is a codimension-two
surface rather than a geodesic. A higher-dimensional brane source
would change the type and the charge of the probe. To retain the same
string source, we instead consider $\mathcal{N}$ radially extended
F1-string probes distributed uniformly over the compact transverse
section

\begin{equation}
\Sigma_{8}=S_{w}^{1}\times S^{3}\times T^{4},
\end{equation}

\noindent at the horizon. Each radial string pierces one of $\mathcal{N}$
equal cells of the horizon section, 

\begin{equation}
\Delta\mathcal{V}_{0}=\frac{\mathcal{V}_{0}}{\mathcal{N}},
\end{equation}

\noindent as illustrated in figure (\ref{fig:Nstring}). The corresponding
probe limit now becomes

\begin{equation}
\epsilon_{\mathrm{probe}}^{\left(10\right)}\sim\frac{G_{N}^{\left(10\right)}\mathcal{N}T_{\mathrm{probe}}}{L^{6}}\ll1.
\end{equation}

Let the transverse coordinates be denoted collectively by $\xi=\left(w,y^{i},\Omega^{\alpha}\right)$,
and introduce the reference measure

\begin{equation}
d\mu_{0}\left(\xi\right)=\sqrt{\bar{\gamma}\left(\Omega\right)}dwd^{4}yd^{3}\Omega.
\end{equation}

\noindent The corresponding reference volume is

\begin{equation}
\mathcal{V}_{0}=L_{w}V_{T^{4}}\Omega_{3},\qquad\Omega_{3}=2\pi^{2}.
\end{equation}

\noindent Here $L_{w}=\int dw$ and $V_{T^{4}}=\int d^{4}y$ denote
coordinate reference volumes. Thus, $\mathcal{V}_{0}$ is independent
of the dynamical string-frame metric and should not be confused with
the proper volume of $\Sigma_{8}$. Define 

\noindent 
\begin{equation}
\Delta_{8}\left(\xi,\xi_{n}\right)\equiv\delta\left(w-w_{n}\right)\delta_{T^{4}}^{\left(4\right)}\left(y-y_{n}\right)\delta_{S^{3}}\left(\Omega,\Omega_{n}\right),
\end{equation}
which is normalized according to

\begin{equation}
\int_{\Sigma_{8}}d\mu_{0}\left(\xi\right)\Delta_{8}\left(\xi,\xi_{n}\right)=1.
\end{equation}

\noindent For the $n$-th radial string, the only non-vanishing component
of the Kalb--Ramond current is

\begin{equation}
j_{\mathrm{F1}}^{0r}=\frac{\sqrt{\bar{\gamma}}}{\sqrt{-g}}\Delta_{8}\left(\xi,\xi_{n}\right).
\end{equation}

\noindent Consequently, the current carried by $\mathcal{N}$ strings
is

\begin{equation}
j_{\mathrm{F1}}^{0r}=\frac{\sqrt{\bar{\gamma}}}{\sqrt{-g}}\sum_{n=1}^{\mathcal{N}}\Delta_{8}\left(\xi,\xi_{n}\right).
\end{equation}

\noindent To define the continuum limit, divide $\Sigma_{8}$ into
$\mathcal{N}$ cells of equal reference volume $\Delta\mathcal{V}_{0}=\mathcal{V}_{0}/\mathcal{N}$
and place one string in each cell. The corresponding Riemann sum obeys

\begin{equation}
\Delta\mathcal{V}_{0}\sum_{n=1}^{\mathcal{N}}\Delta_{8}\left(\xi,\xi_{n}\right)\longrightarrow1,
\end{equation}

\noindent in the distributional sense. Therefore, the smeared current
becomes 

\begin{equation}
j_{\mathrm{F1}}^{0r}=\frac{\sqrt{\bar{\gamma}}}{\sqrt{-g}}\frac{\mathcal{N}}{\mathcal{V}_{0}}.
\end{equation}

\noindent We next restrict the construction to a nine-dimensional
constant-time slice $\Sigma_{t}$. Its volume element factorizes as

\begin{equation}
\sqrt{h}=\sqrt{g_{rr}}\sqrt{g_{\mathcal{M}_{8}}}.
\end{equation}

\noindent For the F1-NS5-P solution, define $Y_{\mathrm{F1}}\left(r\right)\equiv\sqrt{g_{ww}}\sqrt{\det\,g_{S^{3}}}\sqrt{\det\,g_{T^{4}}}=r^{3}H_{1}^{-1/2}\left(r\right)H_{5}^{3/2}\left(r\right)H_{P}^{1/2}\left(r\right)$.
Then

\begin{equation}
\sqrt{g_{\mathcal{M}_{8}}}=Y_{\mathrm{F1}}\left(r\right)\sqrt{\bar{\gamma}},
\end{equation}

\noindent where $H_{P}\left(r\right)\equiv1+K\left(r\right)$. To
obtain a divergenceless vector field on $\Sigma_{t}$, define

\begin{equation}
q_{\mathrm{F1}}^{I}\equiv\frac{\sqrt{-g}}{\sqrt{h}}j_{\mathrm{F1}}^{0I},\qquad I=r,w,y^{i},\Omega^{\alpha}.
\end{equation}

\noindent Using the ADM relation $\sqrt{-g}=N_{\mathrm{ADM}}\sqrt{h}$,
where $N_{\mathrm{ADM}}$ is the lapse, one has

\begin{equation}
q_{\mathrm{F1}}^{I}=N_{\mathrm{ADM}}j_{\mathrm{F1}}^{0I},
\end{equation}

\noindent and

\begin{equation}
\partial_{I}\left(\sqrt{h}q_{\mathrm{F1}}^{I}\right)=\partial_{I}\left(\sqrt{-g}j_{\mathrm{F1}}^{0I}\right)=0.
\end{equation}

\noindent Following our previous construction, the bit-thread flow
can be given by

\begin{equation}
v_{\mathrm{F1}}^{I}=C_{{\rm geom}}^{\mathrm{F1}}q_{\mathrm{F1}}^{I}.
\end{equation}

\noindent Because the full exterior is asymptotically flat rather
than asymptotically AdS, ``bit-thread flow'' here means a divergence-free
entropy flow obeying the local capacity bound, with the horizon section
as the relevant cut. Its boundary-entanglement interpretation is invoked
only after passing to the decoupled AdS$_{3}$ throat in Section 6.
For the non-vanishing radial component,

\begin{equation}
v_{\mathrm{F1}}^{r}\left(r\right)=C_{{\rm geom}}^{\mathrm{F1}}\frac{\sqrt{-g}}{\sqrt{h}}j_{\mathrm{F1}}^{0r}=\frac{C_{{\rm geom}}^{\mathrm{F1}}}{\sqrt{g_{rr}}Y_{\mathrm{F1}}\left(r\right)}\frac{\mathcal{N}}{\mathcal{V}_{0}},
\end{equation}

\noindent and hence

\begin{equation}
\left|v_{\mathrm{F1}}\left(r\right)\right|=\sqrt{h_{IJ}v_{\mathrm{F1}}^{I}v_{\mathrm{F1}}^{J}}=\sqrt{g_{rr}}v_{\mathrm{F1}}^{r}=\frac{C_{{\rm geom}}^{\mathrm{F1}}}{Y_{\mathrm{F1}}\left(r\right)}\frac{\mathcal{N}}{\mathcal{V}_{0}}.
\end{equation}

To determine $C_{{\rm geom}}^{{\rm F1}}$, we start from the standard
bit-thread norm bound, which is naturally formulated in the ten-dimensional
Einstein frame, 
\begin{equation}
\left|v_{{\rm E}}\right|\leq\frac{1}{4G_{N}^{\left(10\right)}}.
\end{equation}

\noindent Here, we temporarily use the subscript ``$\mathrm{E}$''
to denote the Einstein frame, and the subscript ``$\mathrm{string}$''
to denote the string frame. The ten-dimensional Einstein- and string-frame
metrics are related by 
\begin{equation}
g_{\mu\nu}^{{\rm E}}=e^{-\phi/2}g_{\mu\nu}^{{\rm String}},
\end{equation}

\noindent due to $g_{\mu\nu}^{{\rm E}}=\exp\left(-\frac{4\phi}{D-2}\right)g_{\mu\nu}^{{\rm String}}$.
Restricting this relation to a constant-time spatial slice gives 
\begin{equation}
h_{ij}^{{\rm E}}=e^{-\phi/2}h_{ij}^{{\rm String}},
\end{equation}
where $h_{ij}$ is the induced nine-dimensional spatial metric. Since
the spatial slice has nine dimensions, the corresponding volume elements
satisfy 
\begin{equation}
\sqrt{h_{{\rm E}}}=e^{-9\phi/4}\sqrt{h_{{\rm String}}}.
\end{equation}

\noindent A bit-thread flow obeys the divergence-free condition 
\begin{equation}
\nabla_{i}v^{i}=\frac{1}{\sqrt{h}}\partial_{i}\left(\sqrt{h}v^{i}\right)=0.
\end{equation}
Therefore, the same conserved flow written in the two frames must
satisfy 
\begin{equation}
\sqrt{h_{{\rm String}}}v_{{\rm String}}^{i}=\sqrt{h_{{\rm E}}}v_{{\rm E}}^{i}.
\end{equation}
Using the transformation of the spatial volume element, we obtain
\begin{equation}
v_{{\rm String}}^{i}=e^{-9\phi/4}v_{{\rm E}}^{i}.
\end{equation}
This relation ensures that the divergence-free condition is preserved
between the two frames. The entropy flux is evaluated through an eight-dimensional
surface $\Sigma_{8}$ inside the nine-dimensional spatial slice, 
\begin{equation}
\Phi=\int_{\Sigma_{8}}v^{i}n_{i}dA.
\end{equation}
Under the Weyl rescaling, the eight-dimensional area element transforms
as 
\begin{equation}
dA_{{\rm E}}=e^{-2\phi}dA_{{\rm String}},
\end{equation}
while the unit normal covector transforms as 
\begin{equation}
n_{i}^{{\rm E}}=e^{-\phi/4}n_{i}^{{\rm String}}.
\end{equation}
Together with 
\begin{equation}
v_{{\rm String}}^{i}=e^{-9\phi/4}v_{{\rm E}}^{i},
\end{equation}
these relations give 
\begin{equation}
v_{{\rm String}}^{i}n_{i}^{{\rm String}}dA_{{\rm String}}=v_{{\rm E}}^{i}n_{i}^{{\rm E}}dA_{{\rm E}}.
\end{equation}
Thus the physical entropy flux is the same in the Einstein and string
frames. Moreover, the norm of the flow, however, transforms nontrivially.
Since 
\begin{equation}
h_{ij}^{{\rm String}}=e^{\phi/2}h_{ij}^{{\rm E}},
\end{equation}
we find 
\begin{align}
\left|v_{{\rm String}}\right|^{2} & =h_{ij}^{{\rm String}}v_{{\rm String}}^{i}v_{{\rm String}}^{j}\nonumber \\
 & =e^{\phi/2}h_{ij}^{{\rm E}}\left(e^{-9\phi/4}v_{{\rm E}}^{i}\right)\left(e^{-9\phi/4}v_{{\rm E}}^{j}\right)\nonumber \\
 & =e^{-4\phi}\left|v_{{\rm E}}\right|^{2}.
\end{align}
Therefore, 
\begin{equation}
\left|v_{{\rm String}}\right|=e^{-2\phi}\left|v_{{\rm E}}\right|.
\end{equation}

\noindent Applying the Einstein-frame bit-thread bound then gives
\begin{equation}
\left|v_{{\rm String}}\right|\leq\frac{e^{-2\phi}}{4G_{N}^{\left(10\right)}}.
\end{equation}
Hence, for the F1--NS5--P background, the appropriate string-frame
norm bound is 
\begin{equation}
\left|v_{\mathrm{F1}}\left(r\right)\right|\leq\frac{e^{-2\phi\left(r\right)}}{4G_{N}^{\left(10\right)}}.
\end{equation}

\noindent The factor $e^{-2\phi}$ is therefore not an additional
assumption. It follows directly from rewriting the usual Einstein-frame
bit-thread bound in the string frame. It is also consistent with the
string-frame Wald entropy, whose transverse area measure is weighted
by $e^{-2\phi}$. 

We now choose the flow to saturate this bound at the horizon $r=r_{h}$.
This fixes

\begin{equation}
C_{{\rm geom}}^{\mathrm{F1}}=\frac{e^{-2\phi\left(r_{h}\right)}Y_{\mathrm{F1}}\left(r_{h}\right)}{4G_{N}^{\left(10\right)}}\frac{\mathcal{V}_{0}}{\mathcal{N}}.
\end{equation}

\noindent This choice obeys the bound outside the horizon because

\begin{equation}
e^{-2\phi\left(r\right)}Y_{\mathrm{F1}}\left(r\right)=r^{3}\sqrt{H_{1}\left(r\right)H_{5}\left(r\right)H_{P}\left(r\right)}=\sqrt{\left(r^{2}+Q_{1}\right)\left(r^{2}+Q_{5}\right)\left(r^{2}+Q_{P}\right)},
\end{equation}

\noindent is nondecreasing for $r\geq0$. The complete radial flow
is therefore

\begin{equation}
v_{\mathrm{F1}}^{r}\left(r\right)=\frac{e^{-2\phi\left(r_{h}\right)}}{4G_{N}^{\left(10\right)}\sqrt{g_{rr}}}\frac{Y_{\mathrm{F1}}\left(r_{h}\right)}{Y_{\mathrm{F1}}\left(r\right)}.
\end{equation}

\noindent Its horizon value is

\begin{equation}
v_{\mathrm{F1}}^{r}\left(r_{h}\right)=\frac{1}{4G_{N}^{\left(10\right)}}H_{1}\left(r_{h}\right)H_{5}^{-3/2}\left(r_{h}\right).
\end{equation}

\noindent The Bekenstein-Hawking entropy is the flux through the horizon
section. Since $n_{r}=\sqrt{g_{rr}}$,

\begin{equation}
dS=\sqrt{g_{\Sigma_{8}}}dwd^{4}yd^{3}\Omega=Y_{\mathrm{F1}}\left(r\right)\sqrt{\bar{\gamma}}dwd^{4}yd^{3}\Omega,
\end{equation}

\noindent we obtain

\begin{eqnarray}
\Phi_{\mathrm{F1}} & = & \int v_{\mathrm{F1}}^{r}n_{r}dS\nonumber \\
 & = & \frac{L_{w}V_{T^{4}}2\pi^{2}}{4G_{N}^{\left(10\right)}}r_{h}^{3}\sqrt{H_{1}\left(r_{h}\right)H_{5}\left(r_{h}\right)H_{P}\left(r_{h}\right)}=S_{\mathrm{BH}}^{\mathrm{F1}}.
\end{eqnarray}

\noindent For the extremal solution, $r_{h}=0$, and the finite limiting
value is

\begin{equation}
S_{\mathrm{BH}}^{\mathrm{F1}}=\frac{L_{w}V_{T^{4}}2\pi^{2}}{4G_{N}^{\left(10\right)}}\sqrt{Q_{1}Q_{5}Q_{P}}.
\end{equation}

The same result also clarifies the physical meaning of $C_{{\rm geom}}^{\mathrm{F1}}$.
Since the radial flux is conserved, the flux through any constant-$r$
section can be written as

\noindent 
\begin{equation}
C_{{\rm geom}}^{\mathrm{F1}}=\frac{S_{\mathrm{BH}}^{\mathrm{F1}}}{\mathcal{N}}.\label{eq:F1entropycapacity}
\end{equation}
Thus, $C_{{\rm geom}}^{\mathrm{F1}}$ converts the projected string
charge density into entropy flux and measures how much entropy flux
is carried by one unit of string current. In the equivalent tube description,
it is the entropy-flux capacity assigned to each string-charge tube.
The norm bound limits this capacity, and the horizon is the bottleneck
at which its maximal value is reached. 

This interpretation motivates the \emph{boundary-matching conjecture}
introduced below. If the entanglement entropy of a complete asymptotic
boundary factor fixes the conserved integrated entropy flux, then
it also fixes the maximal carrying capacity for the chosen normalization
of the string current. Current conservation then carries the same
total flux from the boundary to the horizon. This provides a possible
information-flow interpretation of the relation among boundary entanglement
entropy, black hole entropy, and the string current.

\section{D1--D5--P with a macroscopic D1 string source}

We now apply the S-duality transformation to the preceding NS--NS
system. In the truncation containing the string-frame metric, the
dilaton, and the R--R two-form $C^{\left(2\right)\prime}$, the bulk
action is

\begin{equation}
S^{\prime}=\frac{1}{2\kappa_{10}^{2}}\int d^{10}x\sqrt{-g^{\prime}}\left[e^{-2\phi^{\prime}}\left(R^{\prime}+4\left(\nabla^{\prime}\phi^{\prime}\right)^{2}\right)-\frac{1}{12}F_{3}^{2}\right]+S_{\mathrm{D1}},\label{eq:D1-D5-source}
\end{equation}

\noindent where $2\kappa_{10}^{2}=\left(2\pi\right)^{7}\alpha^{\prime4}$
and

\noindent 
\begin{equation}
F_{3}=dC^{\left(2\right)\prime},\qquad F_{\mathrm{abc}}=3\partial_{\left[\mathrm{a}\right.}C_{\left.\mathrm{bc}\right]}^{\left(2\right)\prime},
\end{equation}
The consistent truncation used here sets

\begin{equation}
B^{\prime}=0,\qquad H^{\prime}=0,\qquad C_{0}^{\prime}=0,\qquad F_{1}=0,\qquad F_{5}=0.
\end{equation}

Now, we derive the form of the D1 source action $S_{\mathrm{D1}}$
by applying S-duality to the original F1 source action $S_{\mathrm{F1}}$.
The S-transformation acts as

\begin{equation}
\phi^{\prime}=-\phi,\qquad g_{\mathrm{ab}}^{\prime}=e^{-\phi}g_{\mathrm{ab}},\qquad C_{\mathrm{ab}}^{\left(2\right)\prime}=B_{\mathrm{ab}},
\end{equation}

\noindent or, equivalently

\begin{equation}
g_{\mathrm{ab}}=e^{-\phi^{\prime}}g_{\mathrm{ab}}^{\prime}.
\end{equation}

\noindent Substitution into the F1 source action therefore produces
the factor $e^{-\phi^{\prime}}$ in the kinetic term. The complete
D1-brane action also contains a worldvolume $U\left(1\right)$ gauge
field $A_{m}$ \cite{Bergshoeff:1996tu,Schmidhuber:1996fy}. In the
present truncation, however, we consider a pure D1 source carrying
no additional F1 charge. We may therefore consistently choose the
zero electric flux sector. The full DBI--Wess--Zumino action then
reduces to the following Polyakov-type form:

\begin{equation}
S_{\mathrm{D1}}=-\frac{1}{4\pi\alpha^{\prime}}\int d^{2}\sigma\left(e^{-\phi^{\prime}}\sqrt{-\gamma}\gamma^{mn}\partial_{m}X^{\mathrm{a}}\partial_{n}X^{\mathrm{b}}g_{\mathrm{ab}}^{\prime}+\epsilon^{mn}\partial_{m}X^{\mathrm{a}}\partial_{n}X^{\mathrm{b}}C_{\mathrm{ab}}^{\left(2\right)\prime}\right).
\end{equation}

\noindent Thus, the gauge-field-free action used here should be understood
as the pure-D1, zero worldvolume flux sector of the complete D1-brane
action. It describes a macroscopic D1 source that couples electrically
to $C^{\left(2\right)\prime}$. Because the D1 kinetic term depends
explicitly on $\phi^{\prime}$, the dilaton equation now contains
a localized source. This is an important difference from the F1 case.
The final equations of motion of the action (\ref{eq:D1-D5-source})
are given by

\begin{eqnarray}
R^{\prime\mathrm{ab}}+2\nabla^{\prime\mathrm{a}}\nabla^{\prime\mathrm{b}}\phi^{\prime}-\frac{1}{4}e^{2\phi^{\prime}}\left(F^{\mathrm{acd}}F_{\:\mathrm{cd}}^{\mathrm{b}}-\frac{1}{6}g^{\prime\mathrm{ab}}F^{\mathrm{cde}}F_{\mathrm{cde}}\right) & = & \kappa_{10}^{2}\left(T_{\mathrm{D1}}^{\mathrm{ab}}-\frac{1}{4}g^{\prime\mathrm{ab}}T_{\mathrm{D1}}\right),\nonumber \\
\nabla_{\mathrm{a}}^{\prime}F^{\mathrm{abc}} & = & \frac{\kappa_{10}^{2}}{\pi\alpha^{\prime}}j_{\mathrm{D1}}^{\mathrm{bc}}\nonumber \\
4\nabla^{\prime2}\phi^{\prime}-4\left(\nabla^{\prime}\phi^{\prime}\right)^{2}+R^{\prime} & = & -\frac{\kappa_{10}^{2}}{2}T_{\mathrm{D1}},
\end{eqnarray}

\noindent with the D1 stress tensor and R--R current

\begin{eqnarray}
T_{\mathrm{D1}}^{\mathrm{ab}} & = & -\frac{e^{2\phi^{\prime}}}{2\pi\alpha^{\prime}\sqrt{-g^{\prime}}}\int d^{2}\sigma e^{-\phi^{\prime}\left(X\right)}\sqrt{-\gamma}\gamma^{mn}\partial_{m}X^{\mathrm{a}}\partial_{n}X^{\mathrm{b}}\delta^{\left(10\right)}\left(x-X\left(\tau,\sigma\right)\right),\nonumber \\
j_{\mathrm{D1}}^{\mathrm{ab}} & = & \frac{1}{\sqrt{-g^{\prime}}}\int d^{2}\sigma\left(\frac{\partial X^{\mathrm{a}}}{\partial\tau}\frac{\partial X^{\mathrm{b}}}{\partial\sigma}-\frac{\partial X^{\mathrm{b}}}{\partial\tau}\frac{\partial X^{\mathrm{a}}}{\partial\sigma}\right)\delta^{\left(10\right)}\left(x-X\left(\tau,\sigma\right)\right),\nonumber \\
T_{\mathrm{D1}} & = & g_{\mathrm{ab}}^{\prime}T_{\mathrm{D1}}^{\mathrm{ab}}.
\end{eqnarray}

\noindent As in the F1-NS5-P frame, we also work in the probe approximation:

\begin{equation}
\epsilon_{{\rm probe}}\equiv\frac{G_{N}^{\left(10\right)}T_{{\rm probe}}}{L^{6}}\ll1.
\end{equation}

\noindent To leading order, the background is therefore the source-free
D1--D5--P solution

\begin{eqnarray}
dS_{\mathrm{D1-D5-P}}^{2} & = & \frac{1}{\sqrt{H_{1}H_{5}}}\left[-dt^{2}+dw^{2}+K\left(dt+dw\right)^{2}\right]+\sqrt{H_{1}H_{5}}\left(dr^{2}+r^{2}d\Omega_{3}^{2}\right)+\sqrt{\frac{H_{1}}{H_{5}}}\delta_{ij}dy^{i}dy^{j},\nonumber \\
e^{2\phi^{\prime}} & = & \frac{H_{1}}{H_{5}},\qquad C^{\left(2\right)\prime}=\left(1-H_{1}^{-1}\right)dt\wedge dw+2Q_{5}\omega_{2},
\end{eqnarray}

\noindent where $d\omega_{2}=\omega_{3}$, and

\begin{equation}
H_{1}\left(r\right)=1+\frac{Q_{1}}{r^{2}},\qquad H_{5}\left(r\right)=1+\frac{Q_{5}}{r^{2}},\qquad K\left(r\right)=\frac{Q_{P}}{r^{2}}.
\end{equation}

\noindent The D1-frame metric is conformally related to the F1-frame
metric in its $\left(t,w\right)$ sector, so the ADM shift is unchanged:
$\beta^{\prime w}=K/\left(1+K\right)$. Accordingly, we use the same
slice-adapted embedding as in Eq. (\ref{eq:F1sliceembedding}), rather
than a fixed-$w$ static embedding:

\begin{equation}
X^{t}\left(\tau,\sigma\right)=\tau,\qquad X^{r}\left(\tau,\sigma\right)=\sigma,\qquad X^{w}\left(\tau,\sigma\right)=w_{0}-\frac{K\left(\sigma\right)}{1+K\left(\sigma\right)}\tau,\qquad X^{y^{i}}=y_{0}^{i},\qquad X^{\Omega^{\alpha}}=\Omega_{0}^{\alpha}.
\end{equation}

\noindent At $\tau=0$, the temporal tangent is parallel to the D1-frame
ADM normal and the spatial tangent is radial. In particular,

\begin{equation}
\gamma_{\tau\tau}^{\prime\,\mathrm{ind}}=-\frac{1}{\sqrt{H_{1}H_{5}}\left(1+K\right)},\qquad\gamma_{\tau\sigma}^{\prime\,\mathrm{ind}}=0,\qquad\gamma_{\sigma\sigma}^{\prime\,\mathrm{ind}}=\sqrt{H_{1}H_{5}},
\end{equation}

\noindent and therefore $\det\gamma^{\prime\,\mathrm{ind}}=-1/\left(1+K\right)<0$
at every exterior point. As in the F1 frame, $j_{\mathrm{D1}}^{0w}=0$
on this slice, while the temporal-spatial current component relevant
to the radial projected density is

\begin{equation}
j_{\mathrm{D1}}^{0r}=\frac{\sqrt{\bar{\gamma}}}{\sqrt{-g^{\prime}}}\delta\left(w-w_{0}\right)\delta_{T^{4}}^{\left(4\right)}\left(y-y_{0}\right)\delta_{S^{3}}\left(\Omega,\Omega_{0}\right),
\end{equation}

\noindent For $\mathcal{N}$ radial D1 probes uniformly distributed
over $\Sigma_{8}$, the continuum limit gives

\begin{equation}
j_{\mathrm{D1}}^{0r}=\frac{\sqrt{\bar{\gamma}}}{\sqrt{-g^{\prime}}}\frac{\mathcal{N}}{\mathcal{V}_{0}}.
\end{equation}

\noindent The corresponding probe limit is

\begin{equation}
\epsilon_{\mathrm{probe}}^{\left(10\right)}\sim\frac{G_{N}^{\left(10\right)}\mathcal{N}T_{\mathrm{probe}}}{L^{6}}\ll1.
\end{equation}

\noindent Because S-duality acts on the fields rather than on the
coordinate labels, we retain the same collective coordinate $\xi=\left(w,y^{i},\Omega^{\alpha}\right)$,
the same metric-independent reference measure $d\mu_{0}$, and the
same reference volume

\begin{equation}
\mathcal{V}_{0}=L_{w}V_{T^{4}}\Omega_{3},\qquad\Omega_{3}=2\pi^{2},
\end{equation}

\noindent This does not imply that the proper string-frame volumes
are equal. The current of the $n$-th D1 probe is

\begin{equation}
j_{\mathrm{D1}}^{0r}=\frac{\sqrt{\bar{\gamma}}}{\sqrt{-g^{\prime}}}\Delta_{8}\left(\xi,\xi_{n}\right).
\end{equation}

\noindent Thus, the current carried by $\mathcal{N}$ strings is

\begin{equation}
j_{\mathrm{D1}}^{0r}=\frac{\sqrt{\bar{\gamma}}}{\sqrt{-g^{\prime}}}\sum_{n=1}^{\mathcal{N}}\Delta_{8}\left(\xi,\xi_{n}\right),
\end{equation}

\noindent and the distributional limit

\begin{equation}
\Delta\mathcal{V}_{0}\sum_{n=1}^{\mathcal{N}}\Delta_{8}\left(\xi,\xi_{n}\right)\longrightarrow1,
\end{equation}

\noindent reproduces the uniform current above. As in the F1 frame,
this smearing limit is compatible with the unperturbed background
only when the total D1-probe backreaction remains parametrically small.
On a constant-time slice $\Sigma_{t}$, define

\begin{equation}
\sqrt{h^{\prime}}=\sqrt{g_{rr}^{\prime}}\sqrt{g_{\Sigma_{8}}^{\prime}}.
\end{equation}

\noindent Define $Y_{\mathrm{D1}}\left(r\right)\equiv r^{3}H_{1}^{3/2}H_{5}^{-1/2}H_{P}^{1/2}$
for the D1-D5-P solution, so that

\begin{equation}
\sqrt{g_{\Sigma_{8}}^{\prime}}=Y_{\mathrm{D1}}\left(r\right)\sqrt{\bar{\gamma}}.
\end{equation}

\noindent The projected current

\begin{equation}
q_{\mathrm{D1}}^{I}\equiv\frac{\sqrt{-g^{\prime}}}{\sqrt{h^{\prime}}}j_{\mathrm{D1}}^{0I},\qquad I=r,w,y^{i},\Omega^{\alpha},
\end{equation}

\noindent satisfies

\begin{equation}
\partial_{I}\left(\sqrt{h^{\prime}}q_{\mathrm{D1}}^{I}\right)=0.
\end{equation}

\noindent Equivalently, the ADM decomposition gives $\sqrt{-g^{\prime}}=N_{\mathrm{ADM}}^{\prime}\sqrt{h^{\prime}}$,
and hence

\begin{equation}
q_{\mathrm{D1}}^{I}=N_{\mathrm{ADM}}^{\prime}j_{\mathrm{D1}}^{0I}.
\end{equation}

\noindent The corresponding bit-thread field can be given by

\begin{equation}
v_{\mathrm{D1}}^{I}=C_{{\rm geom}}^{\mathrm{D1}}q_{\mathrm{D1}}^{I}.
\end{equation}

\noindent Its radial component is

\begin{equation}
v_{\mathrm{D1}}^{r}\left(r\right)=C_{{\rm geom}}^{\mathrm{D1}}\frac{\sqrt{-g^{\prime}}}{\sqrt{h^{\prime}}}j_{\mathrm{D1}}^{0r}=\frac{C_{{\rm geom}}^{\mathrm{D1}}}{\sqrt{g_{rr}^{\prime}}Y_{\mathrm{D1}}\left(r\right)}\frac{\mathcal{N}}{\mathcal{V}_{0}},
\end{equation}

\noindent and its norm is

\begin{equation}
\left|v_{\mathrm{D1}}\left(r\right)\right|=\sqrt{h_{IJ}^{\prime}v_{\mathrm{D1}}^{I}v_{\mathrm{D1}}^{J}}=\sqrt{g_{rr}^{\prime}}v_{\mathrm{D1}}^{r}=\frac{C_{{\rm geom}}^{\mathrm{D1}}}{Y_{\mathrm{D1}}\left(r\right)}\frac{\mathcal{N}}{\mathcal{V}_{0}}.
\end{equation}

\noindent The dilaton-weighted bit-thread norm bound is

\begin{equation}
\left|v_{\mathrm{D1}}\left(r\right)\right|\leq\frac{e^{-2\phi^{\prime}\left(r\right)}}{4G_{N}^{\left(10\right)}}.
\end{equation}

\noindent Saturation at $r=r_{h}$ fixes the geometric conversion
coefficient

\begin{equation}
C_{{\rm geom}}^{\mathrm{D1}}=\frac{e^{-2\phi^{\prime}\left(r_{h}\right)}Y_{\mathrm{D1}}\left(r_{h}\right)}{4G_{N}^{\left(10\right)}}\frac{\mathcal{V}_{0}}{\mathcal{N}}.
\end{equation}

\noindent The complete radial flow now becomes

\begin{equation}
v_{\mathrm{D1}}^{r}\left(r\right)=\frac{e^{-2\phi^{\prime}\left(r_{h}\right)}}{4G_{N}^{\left(10\right)}\sqrt{g_{rr}^{\prime}}}\frac{Y_{\mathrm{D1}}\left(r_{h}\right)}{Y_{\mathrm{D1}}\left(r\right)}.
\end{equation}

\noindent Its horizon value is

\begin{equation}
v_{\mathrm{D1}}^{r}\left(r_{h}\right)=\frac{1}{4G_{N}^{\left(10\right)}}H_{1}^{-5/4}\left(r_{h}\right)H_{5}^{3/4}\left(r_{h}\right).
\end{equation}

\noindent Using $e^{-2\phi^{\prime}}=H_{5}/H_{1}$, one finds

\begin{equation}
e^{-2\phi^{\prime}\left(r\right)}Y_{\mathrm{D1}}\left(r\right)=r^{3}\sqrt{H_{1}\left(r\right)H_{5}\left(r\right)H_{P}\left(r\right)}=e^{-2\phi\left(r\right)}Y_{\mathrm{F1}}\left(r\right).
\end{equation}

\noindent The entropy flux is therefore

\begin{eqnarray}
\Phi_{\mathrm{D1}} & = & \int v_{\mathrm{D1}}^{r}n_{r}^{\prime}dS\nonumber \\
 & = & \frac{L_{w}V_{T^{4}}2\pi^{2}}{4G_{N}^{\left(10\right)}}r_{h}^{3}\sqrt{H_{1}\left(r_{h}\right)H_{5}\left(r_{h}\right)H_{P}\left(r_{h}\right)}=S_{\mathrm{BH}}^{\mathrm{D1}}.
\end{eqnarray}

\noindent where $n_{r}^{\prime}=\sqrt{g_{rr}^{\prime}}$ and $dS=Y_{\mathrm{D1}}\left(r\right)\sqrt{\bar{\gamma}}dwd^{4}yd^{3}\Omega$.
Thus, the F1--NS5--P and D1--D5--P frames give the same entropy,
as required by S-duality. In the extremal limit $r_{h}=0$,

\begin{equation}
S_{\mathrm{BH}}^{\mathrm{D1}}=\frac{L_{w}V_{T^{4}}2\pi^{2}}{4G_{N}^{\left(10\right)}}\sqrt{Q_{1}Q_{5}Q_{P}}.
\end{equation}

\section{CFT determination of the radial normalization}

In the above construction, the normalization of the D1 radial flow
was fixed by imposing the dilaton-weighted norm bound

\begin{equation}
\left|v_{\mathrm{D1}}\left(r\right)\right|\leq\frac{e^{-2\phi^{\prime}\left(r\right)}}{4G_{N}^{\left(10\right)}},
\end{equation}

\noindent and requiring saturation at the horizon. We denote the coefficient
fixed in that way by $C_{{\rm geom}}^{\mathrm{D1}}$. We now set this
geometric input aside and return to the conserved projected D1 current
with an undetermined conversion coefficient. We ask whether the D1--D5
CFT fixes a second coefficient, denoted by $C_{{\rm CFT}}^{\mathrm{D1}}$,
without using the horizon area, horizon saturation, the dilaton-weighted
norm bound, or the max-flow/min-cut theorem. Only after this microscopic
determination will we compare $C_{{\rm CFT}}^{\mathrm{D1}}$ with
$C_{{\rm geom}}^{\mathrm{D1}}$ and test the resulting vector throughout
the exterior.

Before proceeding, it is important to distinguish the quantized background
charges 

\begin{equation}
n_{1},\qquad n_{5},\qquad n_{P},
\end{equation}

\noindent from the number $\mathcal{N}$ of auxiliary radial source
lines. The background D1 branes wrap the common circle $S_{w}^{1}$
and, together with the D5 branes, determine the central charge of
the D1--D5 CFT. By contrast, the radial source lines used to represent
the flow are oriented along $\left(t,r\right)$ and are not the background
D1 branes whose bound-state degeneracy is counted by the CFT.

Returning to the D1--D5--P configuration, the uniformly smeared
radial current is 

\begin{equation}
j_{\mathrm{D1}}^{0r}=\frac{\sqrt{\bar{\gamma}}}{\sqrt{-g^{\prime}}}\frac{\mathcal{N}}{\mathcal{V}_{0}},
\end{equation}

\noindent where $\mathcal{V}_{0}=L_{w}V_{T^{4}}\Omega_{3}$ and $\Omega_{3}=2\pi^{2}$.
Projecting this current onto a constant-time slice gives 

\begin{equation}
q_{\mathrm{D1}}^{r}=\frac{\sqrt{-g^{\prime}}}{\sqrt{h^{\prime}}}j_{\mathrm{D1}}^{0r}.
\end{equation}

\noindent Writing the induced volume element as 

\begin{equation}
\sqrt{h^{\prime}}=\sqrt{g_{rr}^{\prime}}Y_{\mathrm{D1}}\left(r\right)\sqrt{\bar{\gamma}}.
\end{equation}

\noindent we obtain 

\begin{equation}
q_{{\rm D1}}^{r}\left(r\right)=\frac{1}{\sqrt{g_{rr}^{\prime}}Y_{{\rm D1}}\left(r\right)}\frac{\mathcal{N}}{\mathcal{V}_{0}}.\label{eq:CFTqradial}
\end{equation}

\noindent At this stage the corresponding candidate bit-thread flow
is

\begin{equation}
v_{\mathrm{D1}}^{r}\left(r\right)=C_{{\rm CFT}}^{\mathrm{D1}}q_{\mathrm{D1}}^{r}\left(r\right),\label{eq:vd1}
\end{equation}

\noindent and its norm is 

\begin{equation}
\left|v_{\mathrm{D1}}\left(r\right)\right|=\sqrt{g_{rr}^{\prime}}v_{\mathrm{D1}}^{r}=\frac{C_{{\rm CFT}}^{\mathrm{D1}}}{Y_{\mathrm{D1}}\left(r\right)}\frac{\mathcal{N}}{\mathcal{V}_{0}}.\label{eq:vnormd1}
\end{equation}

\noindent The flux through the constant $r$ surface is therefore

\begin{equation}
\Phi\left(r\right)=\int v^{r}n_{r}dS=C_{{\rm CFT}}^{\mathrm{D1}}\mathcal{N}.
\end{equation}

We next determine the entropy that is to be assigned to this conserved
flow. The CFT lives on the conformal boundary of the $\mathrm{AdS}_{3}$
decoupling region, whose spatial circle is 
\begin{equation}
w\sim w+L_{w},\qquad L_{w}=2\pi R_{w}.\label{eq:CFTcircle}
\end{equation}
Although the proper circumference of an unrenormalized large-radius
cutoff surface diverges, the boundary CFT is defined with respect
to a finite representative of the boundary conformal metric. Its spatial
coordinate length is therefore $L_{w}$, not the divergent proper
circumference of the cutoff surface.

The BPS degeneracy and the extremal horizon entropy are unchanged
by the decoupling limit. Because the radial probe current has no endpoints
between the throat boundary and the horizon, current conservation
allows a normalization fixed in the throat to be extended through
the full asymptotically flat exterior.

For the extremal D1--D5--P state, the left- and right-moving central
charges are 
\begin{equation}
c_{L}=c_{R}=6n_{1}n_{5}.\label{eq:CFTcentralcharges}
\end{equation}
Choosing the orientation for which positive momentum is carried by
the left-moving sector, the Ramond-sector excitation levels are 
\begin{equation}
N_{L}\equiv L_{0}-\frac{c_{L}}{24}=n_{P},\qquad N_{R}\equiv\bar{L}_{0}-\frac{c_{R}}{24}=0.\label{eq:CFTlevels}
\end{equation}
The subtractions by $c_{L,R}/24$ remove the Ramond-sector ground-state
weights. In the Cardy regime, the microscopic entropy is 
\begin{equation}
S_{{\rm CFT}}=2\pi\sqrt{\frac{c_{L}N_{L}}{6}}=2\pi\sqrt{n_{1}n_{5}n_{P}}.\label{eq:Cardy}
\end{equation}

To interpret this microcanonical entropy as the entanglement entropy
of one complete boundary factor, consider the thermofield double (TFD)
formalism \cite{Azeyanagi:2007bj}:

\begin{equation}
|\Psi_{{\rm micro}}\rangle=\frac{1}{\sqrt{d\left(N\right)}}\sum_{a=1}^{d\left(N\right)}|a\rangle_{L}|a\rangle_{R},\qquad S\left(R\right)=\log d\left(N\right)=S_{{\rm CFT}},\label{eq:CFTpurification}
\end{equation}

\noindent where $R$ denotes the entire right D1--D5 CFT. The conjecture
below concerns this complete boundary subsystem; it does not identify
the Cardy entropy with the entropy of a generic fixed spatial interval.

After integrating the smeared source measure over $S^{3}\times T^{4}$,
the homogeneous boundary source measure and the homogeneous thermodynamic
entropy measure may be written as 
\begin{equation}
d\mu_{{\rm D1}}^{\left(Q\right)}=\rho_{Q}dw,\qquad\rho_{Q}=\frac{\mathcal{N}}{L_{w}},\label{eq:CFTchargemeasure}
\end{equation}
and 
\begin{equation}
d\mu_{{\rm CFT}}^{\left(S\right)}=s_{w}dw,\qquad s_{w}=\frac{S_{{\rm CFT}}}{L_{w}}.\label{eq:CFTentropymeasure}
\end{equation}
Here $s_{w}$ is a coarse-grained thermodynamic entropy density of
the Cardy ensemble; it is not the fine-grained entropy density of
an individual pure microstate. The \emph{boundary-matching conjecture}
can equivalently be expressed as 
\begin{equation}
d\mu_{{\rm CFT}}^{\left(S\right)}=C_{{\rm CFT}}^{\mathrm{D1}}d\mu_{{\rm D1}}^{\left(Q\right)}.\label{eq:CFTlocalmatching}
\end{equation}
For the homogeneous configuration this gives 
\begin{equation}
C_{{\rm CFT}}^{\mathrm{D1}}=\frac{s_{w}}{\rho_{Q}}=\frac{S_{{\rm CFT}}}{\mathcal{N}},\label{eq:CFTnormalization}
\end{equation}
and is algebraically equivalent to the global relation 
\begin{equation}
\Phi=C_{{\rm CFT}}^{\mathrm{D1}}\mathcal{N}=S_{{\rm CFT}}.\label{eq:CFTglobalmatching}
\end{equation}
Thus, the local measure language provides a useful physical interpretation
of the \emph{boundary-matching conjecture}, but it does not constitute
an independent derivation of that conjecture. For the fixed normalization,
$S_{{\rm CFT}}/\mathcal{N}$ measures the effective entropy-carrying
capacity of each string-current tube. The boundary entropy fixes the
total amount of entropy to be transported, while the string-current
decomposition specifies how this entropy flux is distributed among
the carriers.

Substituting Eq. (\ref{eq:CFTnormalization}) into Eqs. (\ref{eq:vd1})
and (\ref{eq:vnormd1}) gives 
\begin{equation}
v_{\mathrm{D1}}^{r}\left(r\right)=\frac{S_{{\rm CFT}}}{\sqrt{g_{rr}^{\prime}}Y_{\mathrm{D1}}\left(r\right)\mathcal{V}_{0}},\qquad\left|v_{\mathrm{D1}}\left(r\right)\right|=\frac{S_{{\rm CFT}}}{Y_{\mathrm{D1}}\left(r\right)\mathcal{V}_{0}}.\label{eq:CFTnormalizedflow}
\end{equation}

\noindent The auxiliary number $\mathcal{N}$ has cancelled from the
vector field. We now test the CFT-normalized flow against the expected
ten-dimensional string-frame bound. We use \cite{Mathur:2005zp}
\begin{align}
Q_{1} & =\frac{g_{s}\alpha^{\prime3}}{V_{4}}n_{1}, & Q_{5} & =g_{s}\alpha^{\prime}n_{5},\nonumber \\
Q_{P} & =\frac{g_{s}^{2}\alpha^{\prime4}}{R_{w}^{2}V_{4}}n_{P}, & G_{N}^{\left(10\right)} & =8\pi^{6}g_{s}^{2}\alpha^{\prime4},\label{eq:CFTchargedictionary}
\end{align}

\noindent together with 
\begin{equation}
V_{T^{4}}=\left(2\pi\right)^{4}V_{4},\qquad L_{w}=2\pi R_{w},\qquad\Omega_{3}=2\pi^{2}.\label{eq:CFTvolumeconventions}
\end{equation}
These relations imply 
\begin{align}
\frac{\mathcal{V}_{0}}{4G_{N}^{\left(10\right)}}\sqrt{Q_{1}Q_{5}Q_{P}} & =\frac{\left(2\pi R_{w}\right)\left(2\pi\right)^{4}V_{4}\left(2\pi^{2}\right)}{32\pi^{6}g_{s}^{2}\alpha^{\prime4}}\frac{g_{s}^{2}\alpha^{\prime4}}{R_{w}V_{4}}\sqrt{n_{1}n_{5}n_{P}}\nonumber \\
 & =2\pi\sqrt{n_{1}n_{5}n_{P}}=S_{{\rm CFT}}.\label{eq:CFTchargeidentity}
\end{align}
This identity follows from the microscopic Cardy entropy and the quantized
charge dictionary; no horizon area has been evaluated.

Combining Eq. (\ref{eq:CFTchargeidentity}) with the conjectural normalization
in Eq. (\ref{eq:CFTnormalization}) gives the conditional equality

\begin{equation}
C_{{\rm CFT}}^{\mathrm{D1}}=\frac{\mathcal{V}_{0}}{4G_{N}^{\left(10\right)}\mathcal{N}}\sqrt{Q_{1}Q_{5}Q_{P}}=C_{{\rm geom}}^{\mathrm{F1/D1}}.\label{eq:CFTgeomcomparison}
\end{equation}

\noindent Here $C_{{\rm geom}}^{\mathrm{F1/D1}}$ is shorthand for
the equal F1- and D1-frame coefficients obtained from the geometric
norm bound with matched current normalization. Their agreement with
$C_{{\rm CFT}}^{\mathrm{D1}}$ is therefore a microscopic--geometric
comparison, not an assumption inserted into the boundary-matching
conjecture.

The dilaton factor in the string-frame norm bound is
\begin{equation}
e^{-2\phi^{\prime}\left(r\right)}=\frac{H_{5}}{H_{1}}.\label{eq:CFTshifteddilaton}
\end{equation}
For the D1--D5--P geometry, 
\begin{equation}
Y_{{\rm D1}}\left(r\right)=r^{3}H_{1}^{3/2}H_{5}^{-1/2}H_{P}^{1/2},\label{eq:CFTYD1}
\end{equation}
and hence 
\begin{align}
e^{-2\phi^{\prime}\left(r\right)}Y_{{\rm D1}}\left(r\right) & =r^{3}\sqrt{H_{1}\left(r\right)H_{5}\left(r\right)H_{P}\left(r\right)}\nonumber \\
 & =\sqrt{\left(r^{2}+Q_{1}\right)\left(r^{2}+Q_{5}\right)\left(r^{2}+Q_{P}\right)}.\label{eq:CFTweightedarea}
\end{align}
Combining Eqs. (\ref{eq:CFTnormalizedflow}), (\ref{eq:CFTchargeidentity}),
and (\ref{eq:CFTweightedarea}), we find 
\begin{align}
\frac{\left|v_{{\rm D1}}\left(r\right)\right|}{e^{-2\phi^{\prime}\left(r\right)}/4G_{N}^{\left(10\right)}} & =\frac{\sqrt{Q_{1}Q_{5}Q_{P}}}{e^{-2\phi^{\prime}\left(r\right)}Y_{{\rm D1}}\left(r\right)}\nonumber \\
 & =\left[\frac{Q_{1}Q_{5}Q_{P}}{\left(r^{2}+Q_{1}\right)\left(r^{2}+Q_{5}\right)\left(r^{2}+Q_{P}\right)}\right]^{1/2}\leq1,\label{eq:CFTpointwisebound}
\end{align}
where the last inequality holds for $Q_{1},Q_{5},Q_{P}>0$ and $r\geq0$.
Equivalently, 
\begin{equation}
\left|v_{{\rm D1}}\left(r\right)\right|\leq\frac{e^{-2\phi^{\prime}\left(r\right)}}{4G_{N}^{\left(10\right)}}.\label{eq:CFTfinalbound}
\end{equation}
Equality is approached only in the exterior extremal horizon limit
$r\rightarrow0$.

The logic of this result is simple. The horizon area, the horizon-saturation
condition, the geometric norm bound, and the max-flow/min-cut theorem
are not used to determine $C_{{\rm CFT}}^{\mathrm{D1}}$. The microscopic
D1--D5--P counting first fixes $S_{{\rm CFT}}$, while the \emph{boundary-matching
conjecture} assigns this entropy to the radial string-charge flux.
Conservation of the string charge density then determines the complete
radial component. Based on this conjecture, we find $C_{{\rm CFT}}^{\mathrm{D1}}=C_{{\rm geom}}^{\mathrm{D1}}$,
and the resulting flow satisfies the expected string-frame norm bound
throughout the exterior region, with saturation only at the extremal
horizon. Thus, the agreement of the two coefficients, the full radial
profile, and the horizon saturation are conditional consequences of
the construction rather than assumptions used to determine $C_{{\rm CFT}}^{\mathrm{D1}}$.
This provides a consistency check of the correspondence in the homogeneous
radial sector. This calculation does not establish a universal norm
bound for every divergenceless vector field. In addition, the S-duality
analysis discussed in Section 7 does not enter the determination of
$C_{{\rm CFT}}^{\mathrm{D1}}$ and therefore provides an independent
test of the correspondence.

The equality of the two coefficients also gives them a simple physical
meaning. From the geometric construction,

\begin{equation}
C_{{\rm geom}}^{\mathrm{F1/D1}}=\frac{S_{{\rm BH}}}{\mathcal{N}}.
\end{equation}

\noindent For a specified unit-current convention and tube number
$\mathcal{N}$, the horizon together with the norm bound fixes the
corresponding entropy-carrying capacity. From the CFT construction,

\begin{equation}
C_{{\rm CFT}}^{\mathrm{D1}}=\frac{S_{{\rm CFT}}\left(R\right)}{\mathcal{N}}.
\end{equation}

\noindent Therefore

\noindent 
\begin{equation}
C_{{\rm CFT}}^{\mathrm{D1}}=\frac{S_{{\rm CFT}}\left(R\right)}{\mathcal{N}}=\frac{S_{{\rm BH}}}{\mathcal{N}}=C_{{\rm geom}}^{\mathrm{F1/D1}}.\label{eq:CFTgeomcapacity}
\end{equation}

\noindent The boundary entanglement entropy and the bulk horizon therefore
give the same entropy-carrying capacity for the string-current tubes.
For a fixed discretization into $\mathcal{N}$ tubes, this means that
the boundary and the horizon assign the same capacity to each tube.
Equivalently, once the capacity of one tube is chosen, they give the
same number of tubes needed to carry the total entropy.

Within the \emph{boundary-matching} conjecture, the same total entropy
flux can therefore be fixed either from the boundary CFT or from the
bulk horizon:

\begin{equation}
\Phi=S_{{\rm CFT}}\left(R\right)=S_{{\rm BH}}.
\end{equation}

\noindent Current conservation then ensures that this flux is the
same through every homologous radial section. In this sense, the boundary
entanglement entropy and the black hole entropy provide two descriptions
of the same conserved entropy flux. This gives an information-flow
interpretation of their equality in the homogeneous radial sector
considered here.

\section{S-duality of string charge density and bit threads}

The previous analysis shows that the correspondence between string
charge density and bit threads is not restricted to the NS--NS sector
of Type IIB supergravity. Under S-duality, the F1--NS5--P system
is mapped to the D1--D5--P system, while the Kalb--Ramond charge
carried by the fundamental strings is mapped to the R--R charge carried
by D1-branes. Correspondingly, the bit-thread flow constructed from
the NS--NS string current is mapped to a bit-thread flow constructed
from the R--R current. Although the explicit form of the vector field
depends on the duality frame, its entropy flux (Bekenstein-Hawking
entropy) is invariant.

Let us recall the S-duality transformation:

\begin{equation}
\phi^{\prime}=-\phi,\qquad g_{\mathrm{ab}}^{\prime}=e^{-\phi}g_{\mathrm{ab}},\qquad C_{\mathrm{ab}}^{\left(2\right)\prime}=B_{\mathrm{ab}}^{\left(2\right)},
\end{equation}

\noindent or, equivalently

\begin{equation}
g_{\mathrm{ab}}=e^{-\phi^{\prime}}g_{\mathrm{ab}}^{\prime}.
\end{equation}

\noindent The correct string-frame transformation is $g_{\mathrm{ab}}^{\prime}=e^{-\phi}g_{\mathrm{ab}}$.
This ensures that the Einstein-frame metric $g_{\mathrm{ab}}^{E}=e^{-\phi/2}g_{\mathrm{ab}}$
remains invariant. Meanwhile, the NS--NS and R--R two-form potentials
are exchanged, up to an orientation-dependent sign:

\begin{equation}
B^{\left(2\right)}\longleftrightarrow C^{\left(2\right)\prime},\qquad H^{\left(3\right)}\longleftrightarrow F^{\left(3\right)}.
\end{equation}

\noindent The F1 source action

\begin{equation}
S_{\mathrm{F1}}=-\frac{1}{4\pi\alpha^{\prime}}\int d^{2}\sigma\left(\sqrt{-\gamma}\gamma^{mn}\partial_{m}X^{\mathrm{a}}\partial_{n}X^{\mathrm{b}}g_{\mathrm{ab}}+\epsilon^{mn}\partial_{m}X^{\mathrm{a}}\partial_{n}X^{\mathrm{b}}B_{\mathrm{ab}}\right),
\end{equation}

\noindent is consequently mapped to

\begin{equation}
S_{\mathrm{D1}}=-\frac{1}{4\pi\alpha^{\prime}}\int d^{2}\sigma\left(e^{-\phi^{\prime}}\sqrt{-\gamma}\gamma^{mn}\partial_{m}X^{\mathrm{a}}\partial_{n}X^{\mathrm{b}}g_{\mathrm{ab}}^{\prime}+\epsilon^{mn}\partial_{m}X^{\mathrm{a}}\partial_{n}X^{\mathrm{b}}C_{\mathrm{ab}}^{\left(2\right)\prime}\right).
\end{equation}

\noindent Thus, the fundamental-string current sourcing the Kalb--Ramond
field is mapped to a D1-brane current sourcing the R--R two-form:

\begin{equation}
j_{\mathrm{F1}}^{\mathrm{ab}}\quad\xrightarrow{\quad S\quad}\quad j_{\mathrm{D1}}^{\mathrm{ab}}.
\end{equation}

\noindent The corresponding projected string charge densities are

\begin{equation}
q_{\mathrm{F1}}^{I}=\frac{\sqrt{-g}}{\sqrt{h}}j_{\mathrm{F1}}^{0I},\qquad q_{\mathrm{D1}}^{I}=\frac{\sqrt{-g^{\prime}}}{\sqrt{h^{\prime}}}j_{\mathrm{D1}}^{0I}.
\end{equation}

\noindent Both satisfy the divergenceless condition in their respective
duality frames:

\begin{equation}
\partial_{I}\left(\sqrt{h}q_{\mathrm{F1}}^{I}\right)=0,\qquad\partial_{I}\left(\sqrt{h^{\prime}}q_{\mathrm{D1}}^{I}\right)=0.
\end{equation}

\noindent It follows that S-duality maps the NS--NS realization of
the bit-thread flow to an R--R realization:

\begin{equation}
v_{\mathrm{F1}}^{I}=C_{\mathrm{F1}}q_{\mathrm{F1}}^{I}\quad\xrightarrow{\quad S\quad}\quad v_{\mathrm{D1}}^{I}=C_{\mathrm{D1}}q_{\mathrm{D1}}^{I}.
\end{equation}

\noindent Taking the exterior limit toward the extremal horizon,

\begin{equation}
v_{\mathrm{F1}}^{r}\left(r_{h}\right)=\frac{1}{4G_{N}^{\left(10\right)}}H_{1}\left(r_{h}\right)H_{5}^{-3/2}\left(r_{h}\right),\qquad v_{\mathrm{D1}}^{r}\left(r_{h}\right)=\frac{1}{4G_{N}^{\left(10\right)}}H_{1}^{-5/4}\left(r_{h}\right)H_{5}^{3/4}\left(r_{h}\right).
\end{equation}

\noindent Therefore, the string-frame coordinate representations of
the bit-thread flows are not invariant under S-duality. The metric,
induced spatial measure, dilaton, and coordinate current components
transform. The local vector components remain frame dependent, whereas
the invariant quantity is the integrated entropy flux evaluated using
the appropriate metric and dilaton in each frame. To see it, let us
recall the dilaton-weighted norm bound and entropy flux of F1-NS5-P
and D1-D5-P solutions. Define

\begin{equation}
Y_{\mathrm{F1}}\left(r\right)=r^{3}H_{1}^{-1/2}\left(r\right)H_{5}^{3/2}\left(r\right)H_{P}^{1/2}\left(r\right),\qquad Y_{\mathrm{D1}}\left(r\right)=r^{3}H_{1}^{3/2}H_{5}^{-1/2}H_{P}^{1/2}.
\end{equation}

\noindent The F1 and D1 bit-thread norms are bounded by

\begin{equation}
\left|v_{\mathrm{F1}}\left(r\right)\right|\leq\frac{e^{-2\phi\left(r\right)}}{4G_{N}^{\left(10\right)}},\qquad\left|v_{\mathrm{D1}}\left(r\right)\right|\leq\frac{e^{-2\phi^{\prime}\left(r\right)}}{4G_{N}^{\left(10\right)}}.
\end{equation}

\noindent Saturation at the horizon gives

\begin{equation}
C_{\mathrm{F1}}=\frac{e^{-2\phi\left(r_{h}\right)}Y_{\mathrm{F1}}\left(r_{h}\right)}{4G_{N}^{\left(10\right)}}\frac{\mathcal{V}_{0}}{\mathcal{N}},\qquad C_{\mathrm{D1}}=\frac{e^{-2\phi^{\prime}\left(r_{h}\right)}Y_{\mathrm{D1}}\left(r_{h}\right)}{4G_{N}^{\left(10\right)}}\frac{\mathcal{V}_{0}}{\mathcal{N}},
\end{equation}

\noindent and hence

\begin{equation}
v_{\mathrm{F1}}^{r}\left(r\right)=\frac{e^{-2\phi\left(r_{h}\right)}}{4G_{N}^{\left(10\right)}\sqrt{g_{rr}}}\frac{Y_{\mathrm{F1}}\left(r_{h}\right)}{Y_{\mathrm{F1}}\left(r\right)},\qquad v_{\mathrm{D1}}^{r}\left(r\right)=\frac{e^{-2\phi^{\prime}\left(r_{h}\right)}}{4G_{N}^{\left(10\right)}\sqrt{g_{rr}^{\prime}}}\frac{Y_{\mathrm{D1}}\left(r_{h}\right)}{Y_{\mathrm{D1}}\left(r\right)}.
\end{equation}

\noindent Although the two radial vector fields of bit threads are
different, their dilaton-weighted transverse area densities agree:

\begin{equation}
e^{-2\phi\left(r\right)}Y_{\mathrm{F1}}\left(r\right)=r^{3}\sqrt{H_{1}\left(r\right)H_{5}\left(r\right)H_{P}\left(r\right)},\qquad e^{-2\phi^{\prime}\left(r\right)}Y_{\mathrm{D1}}\left(r\right)=r^{3}\sqrt{H_{1}\left(r\right)H_{5}\left(r\right)H_{P}\left(r\right)}.
\end{equation}

\noindent Therefore, we obtain the invariant quantity under S-duality:

\begin{equation}
e^{-2\phi\left(r\right)}Y_{\mathrm{F1}}\left(r\right)=e^{-2\phi^{\prime}\left(r\right)}Y_{\mathrm{D1}}\left(r\right).
\end{equation}

\noindent This identity is the central reason why the two bit-thread
constructions yield the same entropy flux. To see how this identity
ensures the S-duality invariance of the Bekenstein-Hawking entropy,
we recall

\begin{equation}
S_{\mathrm{BH}}^{\mathrm{F1}}=\int v_{\mathrm{F1}}^{r}n_{r}dS,\qquad S_{\mathrm{BH}}^{\mathrm{D1}}=\int v_{\mathrm{D1}}^{r}n_{r}^{\prime}dS,
\end{equation}

\noindent which give

\begin{equation}
S_{\mathrm{BH}}^{\mathrm{F1}}=\frac{e^{-2\phi\left(r_{h}\right)}Y_{\mathrm{F1}}\left(r_{h}\right)}{4G_{N}^{\left(10\right)}}\int\sqrt{\bar{\gamma}}dwd^{4}yd^{3}\Omega,\qquad S_{\mathrm{BH}}^{\mathrm{D1}}=\frac{e^{-2\phi^{\prime}\left(r_{h}\right)}Y_{\mathrm{D1}}\left(r_{h}\right)}{4G_{N}^{\left(10\right)}}\int\sqrt{\bar{\gamma}}dwd^{4}yd^{3}\Omega.
\end{equation}

\noindent Therefore, the identity ensures the Bekenstein-Hawking entropy
is invariant under S-duality

\begin{equation}
S_{\mathrm{BH}}^{\mathrm{F1}}=S_{\mathrm{BH}}^{\mathrm{D1}}.
\end{equation}

\subsubsection*{Physical interpretation}

Now, we wish to summarize several important physical implications
of S-duality for the string charge density/bit threads correspondence:

\vspace*{2.0ex}

\noindent \textbf{1.} It shows that bit threads should not be identified
exclusively with fundamental strings or with the NS--NS Kalb--Ramond
field. In the F1--NS5--P description, the thread density is encoded
in the F1 current and therefore in the electric charge density associated
with $B$. After S-duality, the same entropy flux is encoded in the
D1 current and the R--R two-form $C^{\left(2\right)\prime}$. Thus,
the physical information represented by the threads is S-duality invariant,
whereas the auxiliary source-current representation assigned to this
information is frame dependent. Schematically, the correspondence
transforms as

\begin{equation}
\text{Kalb-Ramond charge density}\quad\longleftrightarrow\quad\text{F1 bit threads},
\end{equation}

\noindent in the NS--NS frame, while in the R--R frame it becomes

\begin{equation}
\text{R-R two-form charge density}\quad\longleftrightarrow\quad\text{D1 bit threads}.
\end{equation}

\noindent S-duality relates these descriptions:

\begin{equation}
\text{F1 charge threads}\quad\xleftrightarrow{\quad S\quad}\quad\text{D1 charge threads}.
\end{equation}

\vspace*{2.0ex}

\noindent \textbf{2.} This result shows the difference between the
part of a bit threads configuration that is invariant under duality
and the part that depends on the duality frame. When we draw the flow
lines of $v_{\mathrm{F1}}^{I}$ and $v_{\mathrm{D1}}^{I}$ in string-frame
coordinates, their local densities or strengths may look different.
However, the largest number of threads that can cross the horizon
remains unchanged. This happens because the changes in the vector
field are balanced by the changes in the metric, the dilaton, the
unit normal vector, and the area element of the transverse space.
Therefore, the quantity that is invariant under duality is not any
single coordinate component of the flow, but its physical entropy
flux:

\begin{equation}
\Phi=\int v^{I}n_{I}dS.
\end{equation}

\vspace*{2.0ex}

\noindent \textbf{3.} This result provides a nontrivial consistency
check of the proposed string charge density/bit threads correspondence.
Bekenstein-Hawking entropy is invariant under exact string dualities.
If the bit threads construction produced different maximal fluxes
in the F1 and D1 frames, it could not represent a physical entropy.
The equality

\begin{equation}
e^{-2\phi\left(r\right)}Y_{\mathrm{F1}}\left(r\right)=e^{-2\phi^{\prime}\left(r\right)}Y_{\mathrm{D1}}\left(r\right),
\end{equation}

\noindent demonstrates explicitly that the proposed correspondence
respects this requirement.

\vspace*{2.0ex}

\noindent \textbf{4.} The D1 description extends the charge-density
interpretation of bit threads beyond the NS--NS sector. The F1 current
couples to the NS--NS two-form $B$, whereas its S-dual D1 current
couples to the R--R two-form $C^{\left(2\right)}$. Although we set
the asymptotic coupling to the self-dual value $g_{s}=1$, the S-duality
transformation remains nontrivial because the local dilaton, string-frame
metric, source action, and string current all transform. The fact
that the F1-NS5-P and D1-D5-P descriptions give the same entropy flux
therefore shows that the correspondence is not tied to a particular
NS--NS description. Instead, it is consistent with the exact S-duality
of Type IIB string theory. This suggests that the string-charge interpretation
of bit threads may extend to a broader duality-covariant formulation.

\vspace*{2.0ex}

\noindent \textbf{5.} The F1 and D1 strings constitute special elements
of the full $SL\left(2,\mathbb{Z}\right)$ multiplet of $\left(p,q\right)$
strings \cite{Schwarz:1995dk,Townsend:1997kr,Bergshoeff:2006gs,BabaeiVelni:2019ptj}.
The present result therefore suggests a more general correspondence
of the form

\begin{equation}
\left(p,q\right)\text{-string charge density}\quad\longleftrightarrow\quad\left(p,q\right)\text{-bit threads flow}.
\end{equation}

\noindent In such a formulation, the NS--NS and R--R currents would
be organized into an $SL\left(2,\mathbb{Z}\right)$ doublet, while
different $\left(p,q\right)$ descriptions would represent different
duality frames of the same underlying entropy flow. Establishing this
extension requires including a nonvanishing axion and the complete
$SL\left(2,\mathbb{Z}\right)$ covariant Type IIB action. Nevertheless,
the equality of the F1 and D1 entropy fluxes provides direct evidence
for this broader structure.

It is worth noting that the present result is interpreted in the probe
approximation. The radial string sources are assumed not to modify
the background geometry at leading order. A fully backreacted analysis
would require solving the coupled supergravity and source equations
and determining whether the charge density flow continues to satisfy
an appropriate max-flow/min-cut principle. Subject to this constraint,
the S-duality result derived here demonstrates that the string charge
density/bit threads correspondence is compatible with one of the fundamental
nonperturbative symmetries of Type IIB string theory.

\section{Discussion and conclusion}

In this paper, we generalized the string charge density/bit threads
correspondence from three-dimensional backgrounds to ten-dimensional
Type IIB string theory. We studied the extremal F1--NS5--P solution
and its S-dual D1--D5--P solution. Their near-horizon geometries
contain a BTZ factor, which gives a direct link to our previous construction
in three dimensions.

In the F1--NS5--P frame, we introduced a set of radial F1 probes
and distributed them uniformly over the eight-dimensional horizon
section. The spatial projection of the F1 current defines a conserved
radial charge flow. After a suitable normalization, this flow can
be identified with the bit threads. Since the calculation is performed
in the string frame, the norm bound contains a dilaton factor. By
requiring the flow to saturate this bound at the horizon, we fixed
its normalization and showed that its entropy flux reproduces the
Bekenstein--Hawking entropy of the F1--NS5--P solution.

We then applied S-duality to the full construction. Under S-duality,
the F1 current coupled to the NS--NS two-form is mapped to a D1 current
coupled to the R--R two-form. The D1 current also defines a conserved
radial flow and gives the Bekenstein--Hawking entropy of the D1--D5--P
solution. The F1 and D1 bit threads are not the same. Their components
change because the string-frame metric, dilaton, spatial measure,
and norm transform under S-duality. However, these changes cancel
in the physical flux through the horizon. Therefore, the F1 and D1
descriptions give the same entropy flux and the same Bekenstein--Hawking
entropy. This shows that the duality invariant quantity is not a component
of the bit threads, but its physical entropy flux.

The correspondence between string charge density and bit threads has
two complementary normalizations. In the bulk, the bit-thread norm
bound fixes $C_{{\rm geom}}$. Under the boundary-matching conjecture,
D1--D5 CFT counting fixes $C_{{\rm CFT}}$. We find

\begin{equation}
C_{{\rm CFT}}=\frac{S_{{\rm CFT}}\left(R\right)}{\mathcal{N}}=\frac{S_{{\rm BH}}}{\mathcal{N}}=C_{{\rm geom}}.
\end{equation}

\noindent Their agreement provides a consistency check between the
microscopic CFT description and the bulk geometry. It also gives the
two coefficients the same physical meaning: $C$ measures the entropy-flux
capacity per unit of the coarse-grained string charge density.

For a fixed normalization of the string-charge tubes, the boundary
CFT and the bulk horizon therefore give the same capacity for each
tube. Equivalently, once the capacity of one tube is fixed, they give
the same number of tubes needed to carry the total entropy. When $\mathcal{N}$
is only a sampling parameter, however, the number of tubes and the
capacity of each tube are not separately physical. The invariant quantity
is

\begin{equation}
C_{{\rm CFT}}\mathcal{N}=S_{{\rm CFT}}\left(R\right)=S_{{\rm BH}}.
\end{equation}

\noindent Within the \emph{boundary-matching conjecture}, this gives
a simple information-flow picture of the equality between entanglement
entropy and black hole entropy: the same conserved maximal string-charge
flux can be fixed either from the boundary state or from saturation
at the horizon. This may provide a useful link between information
transport, horizon entropy, and entanglement entropy, without assigning
an intrinsic entropy to a single F1 or D1 string.

These results provide complementary tests with different logical status.
The geometrically normalized F1 and D1 flows reproduce the known Bekenstein--Hawking
entropy, and their maximal physical flux is independently shown to
be invariant under the exact S-duality of Type IIB string theory.
Conditional on the boundary-matching conjecture, the D1--D5 CFT supplies
a separate microscopic normalization that agrees with the geometric
one and yields the expected pointwise bound. The D1 description also
extends the correspondence beyond perturbative fundamental strings
and the NS--NS sector: within the probe construction, the same entropy
flux can be represented by an auxiliary D1 current coupled to the
R--R two-form.

The present analysis is based on the probe and continuum approximations.
The radial F1 and D1 sources are assumed not to change the background
geometry at leading order. A fully backreacted treatment would require
solving the coupled supergravity and source equations. 

We end by pointing out three possible directions for future work.
\begin{itemize}
\item In this work, we considered an auxiliary F1 source and its S-dual
D1 source. A natural extension is to study them within the $SL\left(2,\mathbb{Z}\right)$
-covariant formulation of Type IIB string theory and to include general
$\left(p,q\right)$-string sources. This extension requires a nonzero
axion, both the NS--NS and R--R two-form fields, and the complete
covariant source action. It may show how different duality frames
describe the same entropy flow and whether the maximal physical flux
remains duality invariant.
\item Our construction extends the string charge density/bit threads correspondence
from three to ten dimensions. It is therefore natural to study its
dimensional reduction to lower-dimensional theories, especially nearly
AdS$_{2}$ gravity and JT gravity. Under dimensional reduction, higher-dimensional
string charges may become lower-dimensional gauge fields or scalar
charges. It would be interesting to determine whether these reduced
charges can still describe entropy flows and whether a similar correspondence
exists in the SYK/nearly AdS$_{2}$ duality.
\item We have studied the roles of open--closed string duality and S-duality
in the string charge density/bit threads correspondence. A natural
next step is to study T-duality. T-duality can map Type IIB theory
to Type IIA theory, but the candidate source-current representation
of the entropy flow depends on the duality direction. For example,
winding and momentum modes may be exchanged, while a D1-brane may
become a D0- or D2-brane. It is therefore necessary to transform the
source current, compactification measure, dilaton-weighted norm bound,
and flow normalization together. Showing that the maximal entropy
flux is preserved under these transformations would provide a further
nontrivial test of the correspondence. 
\end{itemize}
\vspace{5mm}

\noindent {\bf Acknowledgements} 
HW is supported by NSFC Grant No.12105191. SY is supported by NSFC Grant No.12105031 and No.12547101.

\end{document}